\documentclass[a4paper,fleqn,usenatbib]{mnras}

\usepackage{newtxtext,newtxmath}

\usepackage[T1]{fontenc}
\usepackage{ae,aecompl}

\usepackage{graphicx}	
\usepackage{amsmath}	
\usepackage{lineno}

\usepackage{eso-pic}

\AddToShipoutPictureBG*{%
  \AtPageUpperLeft{%
    \hspace{0.75\paperwidth}%
    \raisebox{-3.5\baselineskip}{%
      \makebox[0pt][l]{\textnormal{DES-2020-0567}}
}}}%

\AddToShipoutPictureBG*{%
  \AtPageUpperLeft{%
    \hspace{0.75\paperwidth}%
    \raisebox{-4.5\baselineskip}{%
      \makebox[0pt][l]{\textnormal{FERMILAB-PUB-21-007-AE}}
}}}%

\usepackage{tikz}
\usetikzlibrary{shapes,arrows}

\tikzstyle{line} = [draw, -latex']
\tikzstyle{cloud} = [draw, ellipse, 
 node distance=2cm, minimum height=2em, text centered, text width=2 cm]
 
\usepackage{soul}

\title[DES-Y1 clusters: cross calibration with SPT]{Exploring the contamination of the DES-Y1 Cluster Sample with SPT-SZ selected clusters}

\author[S. Grandis et al.]{S. Grandis,$^{1,2}$\thanks{E-mail: s.grandis@physik.lmu.de}
J. J. Mohr,$^{1,2,3}$
M. Costanzi,$^{4,5}$ 
A. Saro,$^{4,5,6,7}$ 
S. Bocquet,$^{1, 2}$
M. Klein,$^{1,3}$
\newauthor
M. Aguena,$^{8,9}$ S. Allam,$^{10}$ J. Annis,$^{10}$ B. Ansarinejad,$^{11}$ D. Bacon,$^{12}$ E. Bertin,$^{13,14}$
\newauthor
 L. Bleem,$^{15,16}$ D. Brooks,$^{17}$ D.~L.~Burke,$^{18,19}$ A. Carnero Rosel,$^{20,21}$ M.  Carrasco Kind,$^{22,23}$
\newauthor
 J. Carretero,$^{24}$ F.~J.~Castander,$^{25,26}$ A. Choi,$^{27}$ L.~N.~da Costa,$^{9,28}$ J. De Vincente,$^{29}$ S. Desai,$^{30}$ 
\newauthor 
H.T. Diehl,$^{10}$ J.~P.~Dietrich,$^{1}$ P. Doel,$^{17}$ T.~F.~Eifler,$^{31,32}$ S. Everett,$^{33}$ I. Ferrero,$^{33}$ B. Floyd,$^{34}$  
\newauthor
P. Fosalba,$^{25,26}$ J. Frieman,$^{10,16}$ J.~Garc\'ia-Bellido,$^{36}$ E. Gaztanaga,$^{25,26}$ D. Gruen,$^{18,19,37}$
\newauthor
R.~A.~Gruendl,$^{22,23}$ J. Gschwend,$^{9,28}$  N. Gupta,$^{38}$  G. Gutierrez,$^{10}$ S.~R.~Hinton,$^{39}$ 
\newauthor
D.~L.~Hollowood,$^{33}$ K. Honscheid,$^{27,40}$ D.~J.~James,$^{41}$ T. Jeltema,$^{33}$ K. Kuehn,$^{43,44}$
\newauthor
 O. Lahav,$^{17}$ C. Lidman,$^{44,45}$ M. Lima,$^{8,9}$ M.~A.~G.~Maia,$^{9,28}$ M. March,$^{46}$ J.~L.~Marshall,$^{47}$
\newauthor
P.~Melchior,$^{48}$ F. Menanteau,$^{22,23}$ R. Miquel,$^{24,49}$ R. Morgan,$^{50}$ J. Myles,$^{18,19,37}$ R. Ogando,$^{9,28}$
\newauthor
 A. Palmese,$^{10,16}$ F.~Paz-Chinch\'{o}n,$^{23,51}$ A.~A.~Plazas,$^{48}$ C. L. Reichardt,$^{38}$ A.~K.~Romer,$^{52}$ 
\newauthor
E.~Sanchez,$^{29}$ V.~Scarpine,$^{10}$ S.~Serrano,$^{25,26}$ I.~Sevilla-Noarbe,$^{29}$ P. Singh$^{4,5}$, M.~Smith,$^{53}$
\newauthor
 E.~Suchyta,$^{54}$ M.~E.~C.~Swanson,$^{23}$ G.~Tarle,$^{55}$ D.~Thomas,$^{56}$ C.~To,$^{18,19,37}$ J.~Weller,$^{58,3}$ 
\newauthor
R.D.~Wilkinson,$^{52}$ H. Wu$^{57}$ \\
Affiliations after the  conclusions\\
}

\date{Accepted XXX. Received YYY; in original form ZZZ}

\pubyear{2018}

\hypersetup{draft}
\begin{document}
\label{firstpage}
\pagerange{\pageref{firstpage}--\pageref{lastpage}}
\maketitle

\begin{abstract}
We perform a cross validation of the cluster catalog selected by the red-sequence Matched-filter Probabilistic
Percolation algorithm (redMaPPer) in Dark Energy Survey year 1 (DES-Y1) data by matching it with the Sunyaev-Zel'dovich effect (SZE)
selected cluster catalog from the South Pole Telescope SPT-SZ survey. Of the 1005 redMaPPer selected clusters with measured richness $\hat\lambda>40$ in the joint footprint, 207 are confirmed by SPT-SZ. Using the mass information from the SZE signal, we calibrate
the richness--mass relation using a Bayesian cluster population model. We find a mass trend $\lambda\propto M^{B}$
consistent with a linear relation ($B\sim1$), no significant redshift evolution and an intrinsic scatter in richness of $\sigma_{\lambda} = 0.22\pm0.06$. By
considering two error models, we explore the impact of projection effects on the richness--mass modelling, confirming that such effects are not detectable at the current level of systematic uncertainties.
At low richness SPT-SZ confirms fewer redMaPPer clusters than expected. We interpret this richness dependent deficit in confirmed systems as due to the increased presence at low richness of low mass objects not correctly accounted for by our richness-mass scatter model, which we call contaminants. At a richness $\hat \lambda=40$, this population makes up $>$12$\%$ (97.5 
percentile) of the total population. Extrapolating this to a measured richness $\hat \lambda=20$ yields $>$22$\%$ (97.5 
percentile).
With these contamination fractions, the  predicted redMaPPer number counts in different plausible cosmologies are compatible with the measured abundance. The presence of such a population is also a plausible explanation for the different mass trends ($B\sim0.75$) obtained from mass calibration using purely optically selected clusters.
 The mean mass from stacked weak lensing (WL) measurements suggests that these low mass contaminants are galaxy groups with masses   $\sim3$-
$5\times 10^{13} $ M$_\odot$ which are beyond the sensitivity of current SZE and X-ray surveys but a natural target for SPT-3G and eROSITA.

\end{abstract}




\defcitealias{mcclintock19}{McC19}
\defcitealias{saro15}{S15}
\defcitealias{costanzi19}{C19}

\section{Introduction}

Extraction of cosmological information from the number counts of galaxy clusters is critically sensitive to the contamination of the
selected samples and to their completeness as a function of mass \citep{aguena18}. 
Over the last decade, the method of direct mass calibration
has been established as an empirical approach to the modelling of completeness of cluster samples. By constraining the mean 
relation between selection observable and mass, and the scatter around this relation, the thresholds applied in 
selection observable can be transformed into completeness as a function of mass \citep[see for instance][]{melin05, grandisip}.
While systematic uncertainties on this mapping 
are still large, they can be faithfully traced and propagated onto the cosmological constraints via self-consistent and 
simultaneous analysis of the number counts and the mass calibration \citep[e.g.][]{Mantz15, bocquet19, des_y1_cluster}.
 
Concerning the purity of cluster samples, traditionally, the focus was on adjusting cluster selection in such a way as to 
limit contamination. The Sunyaev-Zel'dovich effect \citep[SZE,][]{sunyeavzeldovich72}, for instance, introduces a distinct spectral feature in the cosmic 
microwave background (CMB).  Multifrequency matched filtering of CMB maps \citep{melin06} therefore can provide pure cluster samples \citep{bleem15, planck16_sze, hilton18, bleem19, huang19, hilton20}. In X-rays 
only extended sources are typically considered \citep{vikhlinin98,boehringer01,romer01, boehringer04, pacaud06, clerc14}, limiting the contamination by point-like non-cluster sources. In both of these methods, which rely on the intracluster medium (ICM) for detection, 
contamination is further controlled by optical confirmation, which is also required to determine the clusters redshifts \citep[for the lastest applications, see][]{klein18, klein19, bleem19}. 

In the case of optical cluster selection via over-densities of red galaxies, no additional multi-wavelength data is required 
to determine the selection observable and the redshift \citep[see for instance][]{rykoff14}. At least in principle, every over-density of red galaxies is expected to be associated with a halo of some mass \citep{cohn07, farahi16}. This is due to the fact that every galaxy lives in a halo. The presence of red galaxies thus guarantees the presence of at least one halo, the host halo of the brightest red galaxy. Assigning the most massive of the -- possibly more than one -- halos to the optical structure theoretically ensures a one-to-one mapping between optically-selected clusters and halos. Several effects need to be accounted for, when modelling the mapping between halo mass and observed richness that results from this mapping. Firstly,  the color filter for the red galaxies, which follows the red 
sequence calibrated on spectroscopic data, sweeps a large range of projected distances along the line of sight. This leads 
to significant projection effects from structures surrounding the cluster \citep{cohn07, song12a, costanzi19}. Furthermore, the optically-selected cluster center can be significantly displaced from the actual halo center, leading to a lower measured richnesses. Other important effects are masking of cluster galaxies, and percolation effects \citep[the merging or splitting of optically-selected clusters,][]{garcia19}.

\citet{costanzi19} and \citet{des_y1_cluster} have recently calibrated the impact of these effects on richness estimates for clusters selected by the red-sequence Matched-filter Probabilistic Percolation algorithm \citep[hereafter redMaPPer, ][]{rykoff14} in the Dark Energy Survey\footnote{\url{https://www.darkenergysurvey.org}} 
\citep[DES,][]{DES16} year 1 data by combining properties extracted from real data with simulations. In this work, we further test this scatter model. We quantify the fraction of optically-selected clusters for which the scatter model fails, calling them \textit{contaminants}. Physically speaking, these objects are low mass halos which suffered projection, percolation, mis-centering or masking effects that are larger than expected from the simulations used to calibrate the scatter model. This type of contamination is in stark contrast with the more traditional use of the term `contaminant' in SZE and X-ray cluster searches, where contaminants are random noise fluctuations or mis-classified point sources. In the context of optical cluster finding, contaminants are low mass halos which are not described by the adopted richness--mass modelling.

Empirical constraints on the optical contamination by low mass systems can be obtained by cross matching ICM selected clusters with optically-selected clusters. The matched systems are 
likely higher mass clusters given their multi-wavelength signature. In contrast, low mass contaminants 
would be associated with shallower 
potential wells filled with less and cooler gas, resulting in weaker SZE signals and X-ray emission. In light of this, 
contamination of optically-selected cluster samples can be studied by investigating the X-ray and SZE properties of these objects \citep{rozo14, rozo15, saro15, saro17, farahi19a}. In this work, we 
shall focus on the validation of optically-selected clusters using SZE information. This is motivated by the large overlap of 
the survey footprints of DES and the South Pole Telescope \citep[SPT, ][]{carlstrom11} SZ survey \citep{story13, bleem15}. This study is further 
facilitated by the extensive cosmological and astrophysical work establishing that the empirically calibrated SZE masses 
derived from the cosmological studies \citep{bocquet15, dehaan16, bocquet19} are consistent with the weak lensing signal, the projected phase space density of galaxy members, and the hot gas 
and stellar content of the SPT selected clusters \citep[see for instance][and references therein]{saro15, hennig17, chiu18, bulbul19, capasso19a}. Specifically, the posteriors on the SZ-signal -- mass relation from the weak lensing calibrated cluster number counts by \citet{bocquet19}, which we use in this work, provide reliable mass information with the appropriate systematic uncertainties for all SPT selected clusters.

In this work, we build on previous studies by \citet{saro15} and \citet{bleem19}, which match SPT-SZ (and SPT-ECS) selected 
clusters \citep{bleem15} to clusters selected by redMaPPer. We first present 
the employed cluster samples (Section~\ref{sec:clustersamples}) and the modeling framework (Section~\ref{sec:methods}). 
To this end, we set up the likelihood for each SPT selected cluster to have a given measured richness conditional on the parameters of the 
richness--mass relation. We then use the SZE mass information to infer the most likely richness--mass relation parameters. Then, 
for each redMaPPer selected object, one can compute the probability that the object is confirmed by SPT-SZ. Contamination 
levels can be estimated by comparing these probabilities with the actual occurrence of matches 
(Section~\ref{sec:crosscalibration}). Given considerations about the incompleteness introduced by the optical cleaning 
of the SPT-SZ candidates compared to the redMaPPer selection, these studies are limited to richnesses larger than 40. We then extrapolate the richness-mass relation we derive to lower richnesses, and investigate if its prediction of the redMaPPer 
number counts is consistent with previous measurements 
(Section~\ref{sec:extrapolation}). We also utilize the mean mass from stacked weak lensing measurements to estimate the mean mass of the contaminants. Finally, we discuss our results in comparison with the literature and outline several future 
prospects of the analysis method we present here (Section~\ref{sec:discussion}). Throughout this work we adopt a flat $\Lambda$CDM cosmology with
$H_\text{0} = 70.6 \text{ km s}^{-1}\text{Mpc}^{-1}$ \citep{rigault18}, $\Omega_\text{M}=0.276$ and $\sigma_8=0.781$ 
\citep{bocquet19}, except where otherwise stated. Masses are computed within a radius at which the mass density is 500 times the critical density of the Universe at the redshifts of the clusters.

\section{Cluster samples}\label{sec:clustersamples}

\begin{figure}
	\includegraphics[width=\columnwidth]{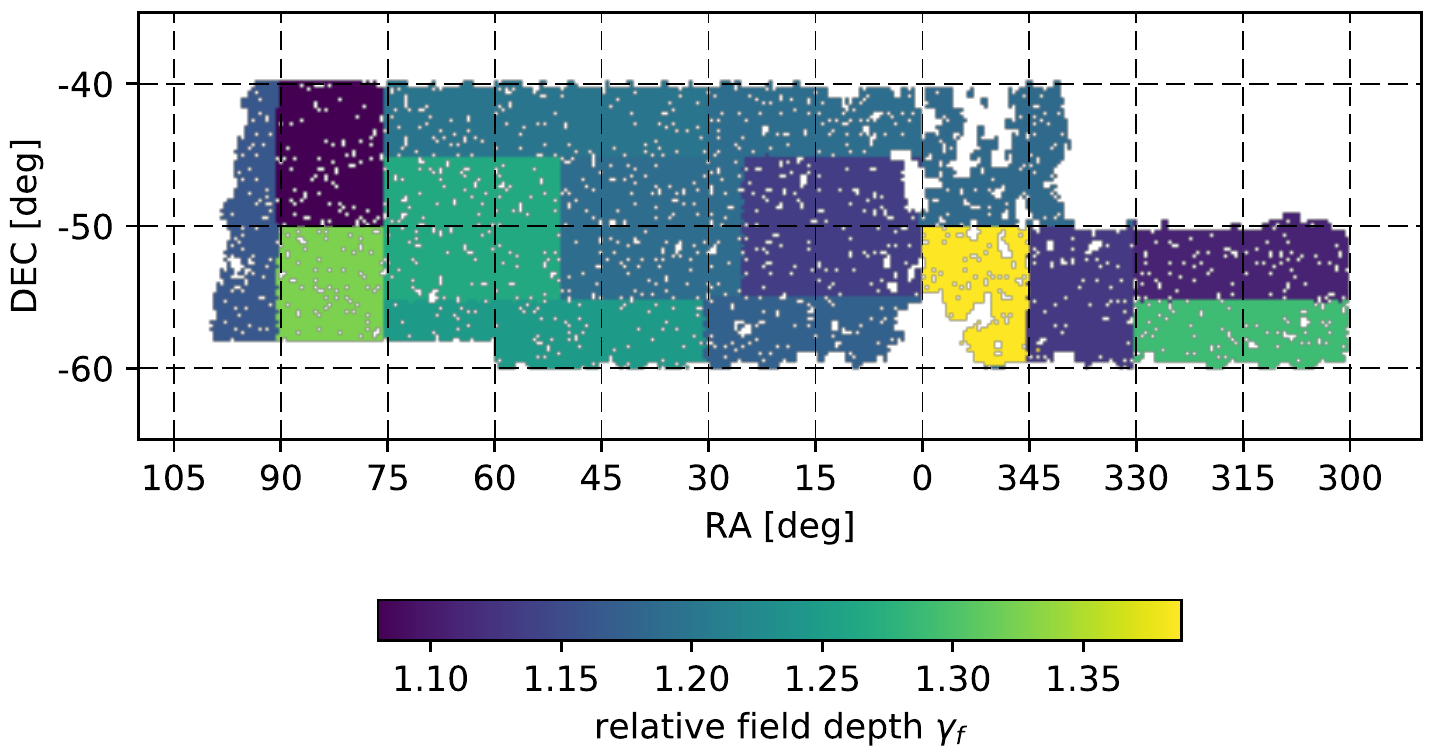}
	\vskip-0.10in
    \caption{Joint SPT-SZ DES-Y1 footprint with colors indicating the relative field depth of SPT-SZ observations as reported in \citet{dehaan16}. The relative field depth allows us to scale the "unbiased significance" -- mass relation to fields of varying depth relative to the definition introduced by \citet{reichardt13}. }
    \label{fig:field_depth_map}
\end{figure}

In this work we investigate the scatter model, contamination fraction and mean scaling relation of the 
optically-selected cluster sample based on the DES-Y1 data  \citep{desy1_data}. These measurements are performed by cross matching and cross 
calibrating the optical sample with clusters selected in the SPT-SZ survey \citep{bleem15}. This limits the analysis to the joint 
footprint of SPT-SZ and DES-Y1, which is shown in Fig.~\ref{fig:field_depth_map} with the relative SPT field depth color 
coded. It comprises an area of $1463$ deg$^2$. In the following, we will touch on the main aspects of the two samples relevant 
to our analysis.

\subsection{Optically-selected samples}

\begin{figure}
	\includegraphics[width=\columnwidth]{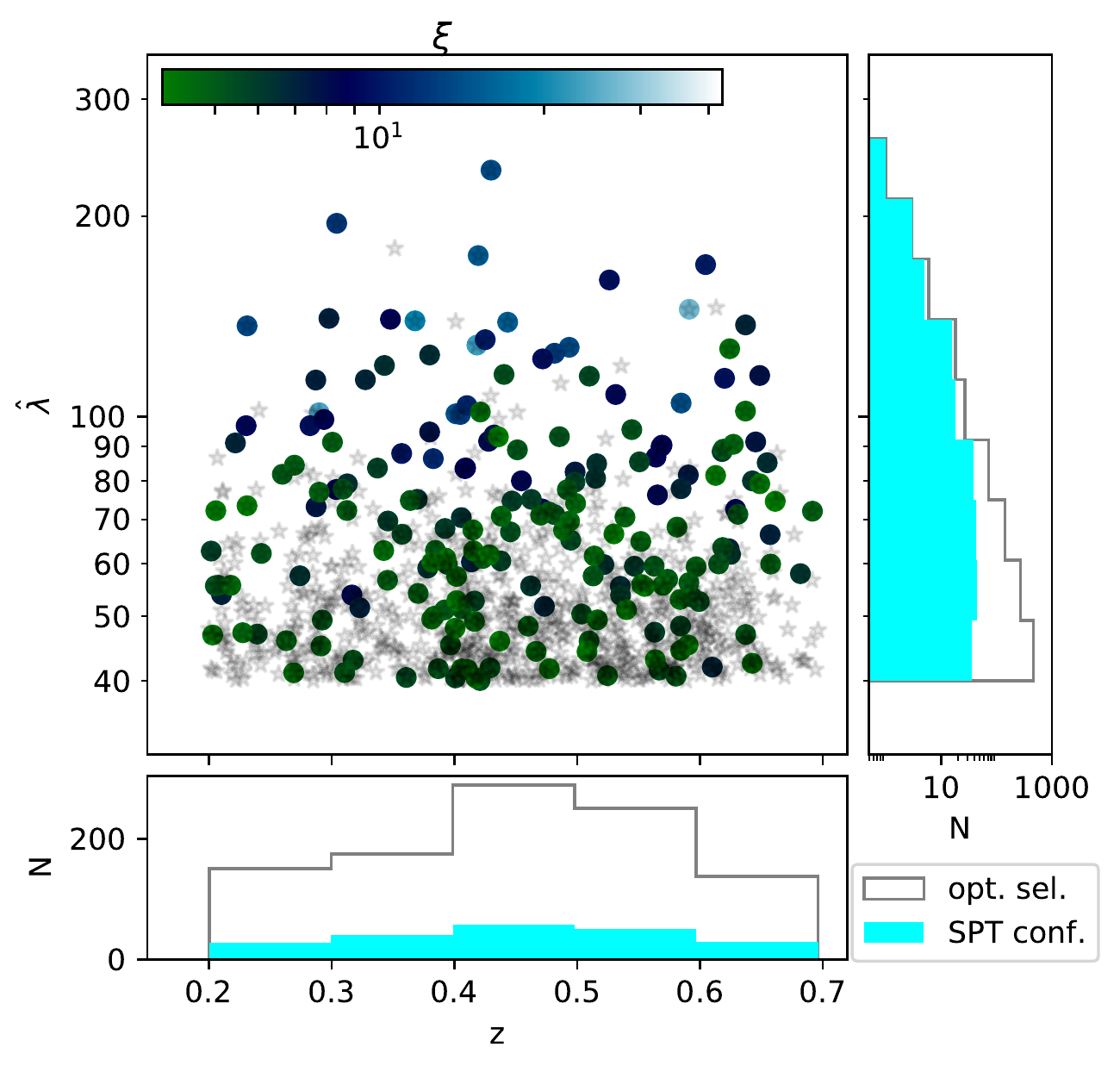}
	\vskip-0.10in
    \caption{\textit{Central panel:} Distribution in measured richness and redshift  of the optically-selected redMaPPer clusters (gray points) with measured richness $
    \hat\lambda>40$ in the joint SPT-SZ--DES-Y1 footprint together with their counterpart SPT-SZ selected clusters, color-coded by their SZE 
    detection significance.
    \textit{Lower panel:} redshift histogram of the  optically-selected redMaPPer clusters with measured richness $
    \hat\lambda>40$ in the joint SPT-SZ--DES-Y1 footprint (opt. sel., in grey) and the redshift histogram of the SPT-SZ confirmed sample (cyan). The fraction of SPT-SZ confirmed objects is constant with redshift. \textit{Right panel:} richness histogram of the optically-selected redMaPPer clusters, and richness histogram of the SPT-SZ confirmed redMaPPer clusters (cyan). The fraction of SPT-SZ confirmed clusters decreases strongly towards lower richness. Quantitatively studying this fraction is the aim of this work.}
    \label{fig:RM_SPT_matched}
\end{figure}

We employ the optically-selected cluster sample extracted from DES-Y1 data \citep{desy1_data} with the redMaPPer algorithm \citep{rykoff14, rykoff16, des_y1_cluster}, which was used for cosmological analyses \citep{des_y1_cluster}. This sample 
provides a measured cluster richness $\hat \lambda^i$ with associated measurement uncertainty $\delta\lambda^i$, and 
photometric redshift $z^i$ for each cluster $i$. The photometric redshifts display percent scatter around spectroscopic redshifts \citep{mcclintock19}. Over the entire DES-Y1 
footprint, when selected by $\hat \lambda>20$, the redMaPPer sample comprises 7066 objects and spans the redshift range $z\in(0.2, 0.7)$. In our cross matching studies, we restrict ourselves to the joint 
DES-Y1 x SPT-SZ footprint and to  $\hat \lambda >40$. This sub-sample consists of 1005 objects, shown as gray points in 
Fig.~\ref{fig:RM_SPT_matched}. For objects in the latter sample we extract the relative SPT field depth $\gamma_f^i$ \citep{dehaan16}. 

\subsection{SPT matched sample}

We match the $\hat \lambda >40$ redMaPPer sample with the SPT-SZ sample selected above SZE signal to noise $\xi>4$ \citep{bleem15}. To 
reduce the contamination by noise fluctuations, we employ the SPT-SZ catalog that was cleaned by the automated cluster 
confirmation and redshift measurement tool MCMF by \citet{kleinip} \citep[see also][for recent applications]{klein18, klein19}, using DES-Y3 
photometric data. MCMF {computes the richness of ICM selected cluster candidates using the ICM signature as a prior for the position and the aperture. The photometric data is filtered using spectroscopically calibrated red-sequence models. Comparison of the candidates richness to the richness distribution along random lines of sight }allows one to select clusters based on the chance of random superposition $f_\text{cont}<0.1$, which enables 
us to produce an uncontaminated SZE selected sample down to signal to noise $\xi=4$. However, in the low SZE signal to noise regime the cut in the chance of random superposition can exclude clusters that would otherwise have passed the ICM selection, introducing optical incompleteness. We show
 in Appendix~\ref{sec:sptcompl} that for objects with 
$\hat \lambda>40$ this incompleteness is always $<2.5\%$. That means that any redMaPPer-object with a measured richness $\hat \lambda>40$ is (almost) certain to make it past the optical cleaning step in the SPT-SZ selection. 
Thus, for the redMaPPer-($\hat \lambda>40$) sample the probability of having an SPT-SZ counterpart is essentially given only by the SZE signal to noise.

We define counterparts in the two samples as those optical and SZE selected clusters lying within a projected radius of 1.0~Mpc 
and having consistent redshifts, i.e. $|z_\text{SPT}-z_\text{RM}| < 0.05 (z_\text{SPT}+z_\text{RM})/2$. We match 207 objects, shown as colored circles in 
Fig.~\ref{fig:RM_SPT_matched}, where the color represents their SZE signal to noise. Two of the SPT clusters (SPT-CLJ0202-5401,  SPT-CLJ0143-4452) are matched to multiple (2) redMaPPer-($\hat \lambda>40$) 
systems. We confirm that in both cases, both redMaPPer objects correspond to a significant detection (i.e. not a random superposition) by MCMF of an optical structure 
along the line of sight to the SPT cluster. The redshift and the MCMF-richness of these objects match the redMaPPer redshifts and richnesses. The SZE signal from these 
objects might have contributions from both clusters along the line of sight. Consequently, the SZE signature is likely biased high. 
However, they do not appear as outliers in the richness -- SZ signal scatter plot.
Given the rarity of these objects, we simply select the redMaPPer objects corresponding to the lowest $f_\text{cont}$ MCMF peak (lowest probability of being a random superposition). 

\section{Methods}\label{sec:methods}

In this section we outline our modelling framework, presenting the hierarchical Bayesian cluster population model used in this work to constrain the richness--mass modelling and the failure fraction of that model from data. We also discuss how the number counts and mean WL masses are used to compare our results to observations.

\subsection{General cluster population model}

\begin{figure}
	\includegraphics[width=\columnwidth]{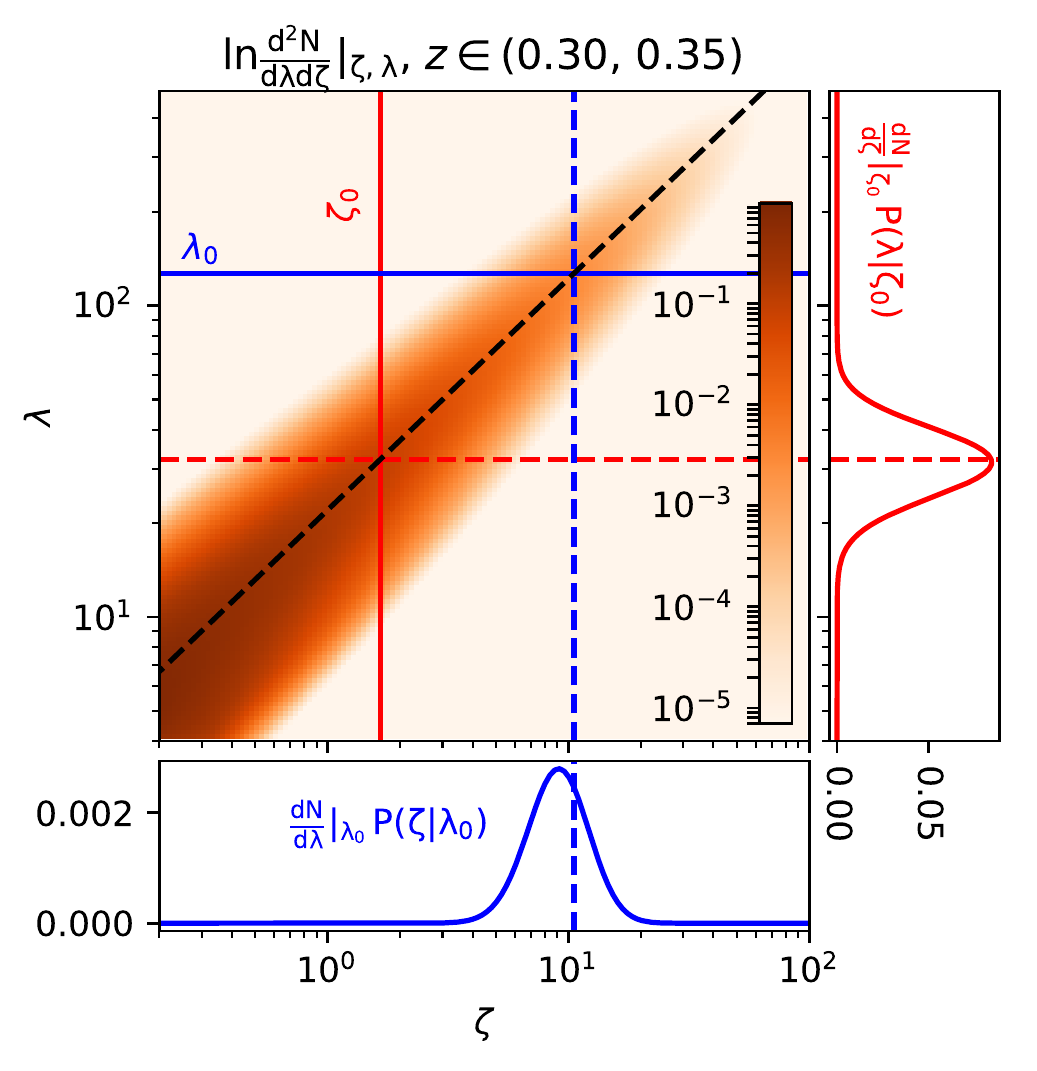}
	\vskip-0.10in
    \caption{Graphic representation of the differential number density in space of intrinsic observables SZE signal $\zeta$ 
    and richness $\lambda$ (central panel). In red is the result of conditioning that distribution on a given value of SZE 
    observable $\zeta_0$ (solid red line), with the resulting conditional probability distribution in richness (right inset). The shape of this distribution represents the 
    richnesses consistent with $\zeta_0$, while the amplitude is the differential number of objects at $\zeta_0$. 
    The dashed line marks the richness obtained by inverting the richness--SZE-signal relation (intersection of solid red line and dashed black line). Note 
    that this richness is larger than the mean expected richness. This effect is caused by Eddington bias, and 
    results from there being more low mass than high mass systems (and thus more systems scattering up than down). The same 
    argument holds when conditioning on a richness $\lambda_0$ in blue and in the lower inset, and explains the offset between the blue curve and blue dashed line in the lower inset when conditioning on richness $\lambda_0$. }
    \label{fig:theory}
\end{figure}

The cluster population model adopted in this work follows closely the model presented in \citet{grandisip}, which builds on work by \citet{bocquet15}. Given the abundance of clusters $\frac{\text{d} N}{\text{d}M}\big|_{M,z}$ as a 
function of mass $M$ and redshift $z$, we express the abundance of clusters in the 
joint space of intrinsic SZE signal to noise $\zeta$ and richness $\lambda$ as 
\begin{equation}
\frac{\text{d}^2 N}{\text{d}\zeta\,\text{d}\lambda}\Big|_{z, \gamma_\text{f}} = \int \text{d}M \, P(\zeta, \lambda|M,z, \gamma_\text{f}) \frac{\text{d} N}{\text{d}M}\Big|_{M,z}
\label{eq:jointabundance}
\end{equation}
where $P(\zeta, \lambda|M,z, \gamma_\text{f}) $ describes the distribution of SZE signal and mean richness for a given mass and redshift.  Because the SZE signature is a detection significance, the expression includes $\gamma_f$, which is the normalized field depth at the cluster position. 
 Furthermore, here we do not yet include measurement noise on the observables.  Thus, $(\zeta, \lambda)$ are {\it intrinsic} observables, as opposed their \textit{measured} counterparts $(\xi, \hat\lambda)$, discussed in Section~\ref{sec:obs_errs}.

A graphic representation 
of the abundance of objects as a function of the two intrinsic observables (with $ \gamma_\text{f}=1$) is shown in the central panel of 
Fig.~\ref{fig:theory}. For these mean relations we adopt the same power law behaviour as outlined in \citet{benson13}, reading 
 \begin{equation}
 \langle \ln \zeta  \rangle =  \ln \gamma_\text{f} A_\text{SZE} +B_\text{SZE} \ln\Big( \frac{M}{M_0} \Big)+ C_\text{SZE} \ln \Big( \frac{E(z)}{E(z_0)} \Big), 
 \label{eq:ximass}
 \end{equation}
for the intrinsic SZE signal, with $M_0=3.0 \times 10^{14} M_{\sun} h^{-1}$ and $z_0=0.6$. 
We define an analogous relation for the mean intrinsic richness, following \citet{saro15}
\begin{equation}
\langle \ln\lambda \rangle = \ln A_\lambda+ B_\lambda \ln \Big( \frac{M}{M_0} \Big) +C_\lambda \ln  \Big( \frac{E(z)}{E(z_0)} \Big).
\label{eq:lambdamass}
\end{equation}

The scatter in SZE signal is modeled as a log-normal distribution with dispersion $\sigma_\text{SZE}$, 
while the scatter in richness has both a log-normal component together with a Poisson contribution $\sigma_{\lambda, \text{tot}}^2 = \sigma^2_\lambda + (\lambda-1)/\lambda^2$. We thus have four free parameters for each relation: an
amplitude $A_{\text{SZE}/\lambda}$, a mass slope $B_{\text{SZE}/\lambda}$, a redshift evolution $C_{\text{SZE}/\lambda}$ and an 
intrinsic scatter $\sigma_{\text{SZE}/\lambda}$. We also introduce the correlation coefficient $\rho$ between the intrinsic scatter in 
SZE signal and the intrinsic scatter in richness as a free parameter of our analysis.

\subsection{Observational errors}\label{sec:obs_errs}

We also account for the observational uncertainties affecting the measured SZE signal and the richness.
For the SZE signal, the measured signal $\xi$ follows the distribution established by \citet{vanderlinde10}, which reads
\begin{equation}
P(\xi|\zeta, \gamma_f) = \frac{1}{\sqrt{2\pi}}\exp\Big\{-\frac{1}{2}\Big( \xi - \sqrt{\zeta^2 + 3} \Big)^2 \Big\}.
\end{equation}

For the richness, we follow two prescriptions. The first follows the method used in \citet{saro15}. 
Together with the measured richness $\hat \lambda^{i}$, the redMaPPer cluster catalog provides an estimate on the error of the 
richness $\delta\lambda^i$ for each entry $i$, which is interpreted as a Gaussian standard deviation, yielding
\begin{equation}\label{eq:errormodellambda}
P_\text{bkg}(\hat \lambda^{i}| \lambda, \delta\lambda^i) = \frac{1}{\sqrt{2\pi (\delta\lambda^i)^2} }\exp\Big\{-\frac{1}{2}\Big(\frac{ \hat\lambda^{i} - \lambda }{\delta\lambda^i}\Big)^2 \Big\}.
\end{equation}
For applications where the average measurement uncertainty as a function of arbitrary measured richness $\hat \lambda$ is 
required, we adopt the extrapolation scheme presented in \citet{grandisip}, appendix A, to estimate $\delta\lambda(\hat\lambda,z)$ 
directly from the catalog. This model accounts only for the photometric uncertainties in the background subtraction. We call this model `background' (`bkg').

A detailed study of projection effects on simulations by \citet{costanzi19}, expanded the prescription above to provide an accurate 
description of the impact of correlated structures, masking and percolation on the mapping between intrinsic and measured richness. This effect is summarized by the
fitted probability density function $P_\text{proj}(\hat\lambda|\lambda,z)$. For the exact definition of this 
function see Eq. 15 in \citet{costanzi19}. This model is called `projection' (`proj').

All analysis steps that follow are performed for both models in an attempt to grasp the impact projection effects might have on our inference.

\subsection{SPT cross calibration and priors}\label{sec:SPTcc}

All objects of the matched redMaPPer-SPT cross-matched sample have a redshift $z^i$, an observed SZE signal $\xi^i$ and a measured richness $\hat\lambda^i$. For each set of scaling relation parameters, we then use  the SZE signal $\xi^i$ 
to predict the expected distribution of intrinsic richnesses $\lambda$, by convolving the joint distribution of intrinsic observables with 
the measurement uncertainty of the SZE signal, i.e.
\begin{equation}
P(\lambda| \xi^i, \gamma_f^i, z^i) \propto \int \text{d} \zeta\, P(\xi^i| \zeta) \frac{\text{d}^2 N}{\text{d}\zeta\,\text{d}\lambda}\Big|_{z^i, \gamma_\text{f}^i}.
\label{eq:lambdagivenxi}
\end{equation}
This equation depends on the scaling relation parameters through the last factor from Eq.~\ref{eq:jointabundance}. As can be 
seen in Fig.~\ref{fig:theory}, for each intrinsic SZE $\xi^i$ this expression defines a range of intrinsic richness at a particular redshift.  Given the typical measured mass-observable scaling relation parameters, higher SZE $\xi^i$ corresponds to higher $\lambda$, and the scatter about both underlying mass--observable relations (Eqs.~\ref{eq:ximass} and \ref{eq:lambdamass}) together with the covariance in this scatter leads to the width in the richness distribution for a given SZE $\xi^i$.

To account for observational uncertainties on richness, Eq.~\ref{eq:lambdagivenxi} can be convolved with the optical error model
\begin{equation}\label{eq:P_measlam_xi}
P(\hat \lambda| \xi^i, \gamma_f^i, z^i) \propto \int \text{d}\lambda\, P(\hat\lambda|\lambda,z)P(\lambda| \xi^i, \gamma_f^i, z^i).
\end{equation}
The proportionality constant is determined by ensuring that the equation above is properly normalized for all possible 
measured richnesses:
$\int_{40}^\infty \text{d}\hat \lambda P(\hat \lambda| \xi^i, \gamma_f^i, z^i) = 1$. 
Note that this normalization cancels any possible dependence on the absolute number of objects, strongly reducing the cosmological dependence in this analysis.

Evaluation of the properly normalized Eq.~\ref{eq:P_measlam_xi} at the measured richness $\hat \lambda^{i}$ gives the likelihood of the measured observables for each cluster in the sample
\begin{equation}\label{eq:crosscalib_lnL}
\ln\mathcal{L}^i = \ln P(\hat \lambda^i| \xi^i, \gamma_f^i, z^i)
\end{equation}
The total log-likelihood then results from summing the log-likelihoods of the individual clusters. 

We proceed with this model by first constraining the richness-mass relation parameters by adopting priors on the SZE-mass scaling relations from a recent, weak lensing informed analysis \citep{bocquet19}. In that work, constraints on
the SZE signal--mass scaling relation parameters are derived by jointly fitting the number counts of cluster sample together with mass calibration information derived from pointed weak lensing follow up measurements. Effectively, we adopt the recent SPT-SZ analysis results (with cosmological constraints in good agreement with other probes) and ask: given the adopted form of the richness-mass relation (Eq.~\ref{eq:lambdamass}), what parameters are required for consistent SZE signals and richnesses of the cross-matched clusters?

Specifically, the priors on the SZE scaling relation parameters in the baseline 
$\nu$-$\Lambda$CDM model read $A_\text{SZE}=5.24\pm0.85$, $B_\text{SZE}=1.534\pm0.100$, $C_\text{SZE}=0.465\pm0.407$, and 
$\sigma_\text{SZE}=0.161\pm0.080$ (symmetrized versions of the constraints reported by \citet{bocquet19}, Table 3, 2nd column `$\nu\Lambda$CDM, SPTcl'). These priors encode the systematic mass
uncertainty on the SZE derived masses. We like to 
stress here that these masses do not assume hydrostatic equilibrium, 
but are instead empirically calibrated using number 
counts and weak lensing information. When predicting the redMaPPer number counts we also use $\Omega_\text{M}=0.276\pm0.047$ and 
$S_8 = \sigma_8 (\Omega_\text{M}/0.3)^{0.2}=0.766\pm0.025$ \citep{bocquet19} to 
properly account for uncertainties in these cosmological parameters. All these priors are modelled as Gaussian distributions in the likelihood inference.

\subsection{Constraining contamination}\label{sec:outliers_and_purity}

\begin{figure}
    \resizebox {\columnwidth} {!} {  
        \begin{tikzpicture}[scale=1, node distance = 1cm, auto]
        
            \node [cloud, ] (c1) {redMaPPer object $\hat \lambda^{i}$, $z^{i}$};
            \node [cloud, below of=c1, left of=c1] (c1a) {is contaminant};
            \node [cloud, below of=c1, right of=c1] (c1b) {is cluster};
            \node [cloud, below of=c1a, left of=c1b] (c2) {is not SPT det.};
            \node [cloud, below of=c1a, right of=c1b] (c3) {is SPT det.};
            
            \path [line] (c1) -- node [anchor=east, xshift=-0.3cm] {$\pi_\text{c}$} (c1a);
            \path [line] (c1) -- node [anchor=west, xshift=0.1cm] {$1-\pi_\text{c}$} (c1b);
            
            \path [line] (c1b) -- node [anchor=east, xshift=-0.1cm] {$1-p^{i}$} (c2);
            \path [line] (c1b) -- node [anchor=west, xshift=0.1cm]{$p^{i}$} (c3);
            
            
            
            \path [line] (c1a) |- node [anchor=west, yshift=1.8cm, xshift=-0.38cm] {1} (c2);

            \node [below of=c2, node distance=0.8cm] {$p(\text{!SPT}|\hat \lambda^{i}, z^{i})$};
            \node [below of=c3, node distance=0.8cm] {$p(\text{SPT}|\hat \lambda^{i}, z^{i})$};
        \end{tikzpicture}
    }
    \vspace{-0.2in}
    \caption{Probability tree descrIbing the different possibilities for a redMaPPer object with measured richness $\hat \lambda^{i}$ and redshift  $z^{i}$ to be detected by SPT $p(\text{SPT}|\hat \lambda^{i}, z^{i})$ and not being detected $p(\text{!SPT}|\hat \lambda^{i}, z^{i})$. These probabilities depend not only on the raw detection probabilities $p^{i}$ as obtained from the observable--mass relations and the selection function, but also on the contamination fraction $\pi_\text{c}$, which models the fraction of objects for which our richness-mass scatter model fails. }\label{fig:sptdetrm_tree}
\end{figure}
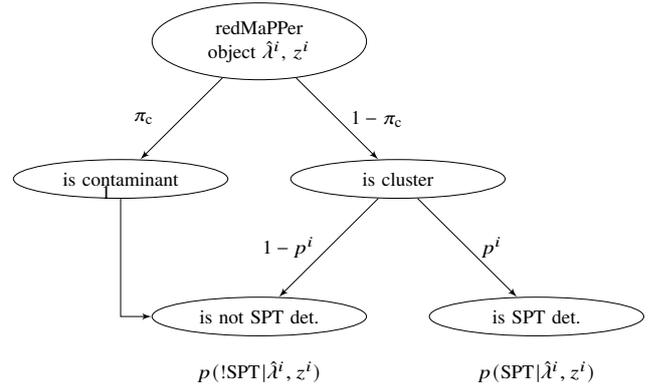

The analysis outlined in the previous section can only be performed on a matched SPT-redMaPPer sample, as for any cluster both a 
measurement of the richness and the SZE signal are required. Considering instead the entire redMaPPer sample above $\hat \lambda>40$, 
we can view being matched or not being matched by SPT as a boolean measurement. We can also seek to predict the outcome of this event for each single redMaPPer cluster $i$ based on the observed richness 
$\hat\lambda^i$, the redshift $z^i$, and the values of the scaling relation parameters. Indeed, given a $\hat\lambda^i$, we 
can predict the probability for a redMaPPer cluster to have a given intrinsic SZE signal by computing 
\begin{equation}\label{eq:P_zeta_g_lambda}
P(\zeta| \hat\lambda^i, z^i, \gamma_\text{f}^i) \propto \int \text{d} \lambda\, P(\hat\lambda^i| \lambda, z^i) \frac{\text{d}^2 N}{\text{d}\zeta\,\text{d}\lambda}\Big|_{z^i, \gamma_\text{f}^i}, 
\end{equation}
where in this case the proportionality constant is set by $\int \text{d}\zeta\,P(\zeta| \hat\lambda^i, z^i, \gamma_\text{f}^i)=1$, as no 
selection of SZE properties was performed. For a graphical representation of this equation, see Fig.~\ref{fig:theory}, where we highlight how conditioning on a given richness ($\lambda_0$ denoted as blue horizontal line) selects a range of intrinsic SZE signal to noises (bottom panel), based on the joint abundance. Note that in this case, too, the normalization cancels the dependence on the absolute number of objects, strongly reducing the cosmological dependence in this analysis.

We can compute the probability $p^i$ that the 
cluster will have a measured SZE signal $\xi>4$ as
\begin{equation}
p^i = \int \text{d}\zeta \, P(\xi>4|\zeta)P(\zeta| \hat\lambda^i, z^i, \gamma_\text{f}^i)
\label{eq:confirmationprobability}
\end{equation}
This is referred to hereafter as the confirmation probability.

In this simple example, the likelihood of matching a system is given by the probability of being confirmed $p^i$, as we remind the 
reader that in a Bayesian context the likelihood is defined as the probability of the data given the model -- albeit that in this case the datum is a boelan (being or not being matched). Similarly, the 
likelihood of a not matched system is given by the probability of not being confirmed, that is $1-p^i$.

As in \citet{grandisip}, we extend this formalism to investigate different properties of the selection function.  
Following the probability tree shown in Fig.~\ref{fig:sptdetrm_tree} from top to bottom,  we first entertain the 
possibility that the redMaPPer object is not described by our richness--mass scatter model, and thus is a contaminant. This is modeled by the contamination fraction $\pi_\text{c}$. Every redMaPPer object has only a 
chance $1-\pi_\text{c}$ to be \text{well described by of richness--mass scatter model} and, conversely, a chance $\pi_\text{c}$ of being a contaminant. If the object is a cluster, the detection probability $p^i$ affects whether it is detected by SPT or not.

Following the probability tree and adding up the weight of all the branches leading to a redMaPPer cluster ending as a non detection (represented by the notation "!SPT"), the likelihood is 
\begin{equation}
p(\text{!SPT}|\hat \lambda^{i}, z^{i}) = \pi_\text{c}+ (1-\pi_\text{c})(1-p^i).
\label{eq:nondetect}
\end{equation}
For matched objects (denoted as "SPT"), the likelihood is 
\begin{equation}
p(\text{SPT}|\hat \lambda^{i}, z^{i}) = (1-\pi_\text{c})p^i.
\label{eq:detect}
\end{equation}
The total likelihood for the redMaPPer sample 
can then be obtained by summing the log-likelihood of the individual clusters based on whether they have been detected by SPT or not, 
reading
\begin{eqnarray}
\label{eq:probs_like}
\ln \mathcal{L} &=&  \sum_{i \in \text{matched}} \ln p(\text{SPT}|\hat \lambda^{i}, z^{i}) \\
& & + \sum_{i \in !\text{matched}} \ln p(\text{!SPT}|\hat \lambda^{i}, z^{i}).\nonumber
\end{eqnarray}
This likelihood depends on the scaling relation parameters as well as on the contamination fraction. 

We employ this model to investigate the case of
a richness dependent contamination (abbreviated "cont" in the following), where the contamination probability is modeled as 
\begin{equation}\label{eq:picont_lambda}
\pi_\text{c}(\hat \lambda) = \frac{A(\hat \lambda)}{1+A(\hat \lambda)}\text{ with } A(\hat \lambda) = A_\text{c} \Big(\frac{\hat \lambda}{45}\Big)^{B_\text{c}}.
\end{equation}
In this parametrisation $ A_\text{c}>0$ and arbitrary  $ B_\text{c}$ lead to values of $0<\pi_\text{c}(\hat \lambda)<1$ for any value of richness $\hat \lambda$.  $ A_\text{c}$ and $ B_\text{c}$  are free parameters of our analysis in this case.

\subsection{Model Predictions of Observables}

After having determined the richness--mass scaling relation parameters, we employ the different posteriors on these parameters
to predict several quantities which we compare with the data: 1) the fraction of SPT detected redMaPPer clusters as a function of redshift and measured richness, 2) the mean mass of clusters in redshift and richness bins, and 3) the number of redMaPPer clusters in redshift and measured richness bins.  These are each described in more detail below.

\subsubsection{SPT confirmation fraction in richness/redshift bin}
\label{sec:confirmationfraction}

The probability of a cluster of intrinsic richness $\lambda$ and redshift $z$ falling into an observed richness and redshift bin $j$, defined by $\hat\lambda_{-}^j<\hat\lambda<\hat\lambda_{+}^j$ and $z_{-}^j<z<z_{+}^j$, is
\begin{equation}
\begin{split}
P(j|\lambda, z) &=  \chi_{z_{-}^j}^{z_{+}^j}(z)\int_{\hat\lambda_{-}^j}^{\hat\lambda_{+}^j} \text{d}\hat \lambda\, P(\hat\lambda|\lambda, z) \\
& = \chi_{z_{-}^j}^{z_{+}^j}(z) \Big[ P(\hat\lambda>\hat\lambda_{-}^j|\lambda, z)-P(\hat\lambda>\hat\lambda_{+}^j|\lambda, z) \Big].
\end{split}
\end{equation}
Here, $ \chi_{z_{-}^j}^{z_{+}^j}(z) =1$ for $z_{-}^j<z<z_{+}^j$, and $0$ elsewhere.
Note that in the limit of infinitely small bins around a measured richness $\hat\lambda^i$ and redshift $z^i$, that is $\hat\lambda_{+/-}^j \rightarrow \hat\lambda^i$ and $z_{+/-}^j \rightarrow z^i$, this equation tends towards the error model for a single cluster (Eq.~\ref{eq:errormodellambda}).

The expected distribution of intrinsic SZE signal $\zeta$ in the bin then follows closely the expression for a single cluster given 
in Eq.~\ref{eq:P_zeta_g_lambda},
\begin{equation}
P(\zeta| j, \gamma_\text{f}) \propto  \int \text{d}z \int \text{d} \lambda\,P(j|\lambda, z) \frac{\text{d}^2 N}{\text{d}\zeta\,\text{d}\lambda}\Big|_{z, \gamma_\text{f}},
\end{equation}
where the normalization is given by the condition $\int\text{d}\zeta\,P(\zeta| j, \gamma_\text{f})=1$. As above, this normalization makes this expression lose much of its dependence on cosmology.

The fraction of the redMaPPer clusters in bin $j$ that are then also in the SPT sample, $f(\text{SPT}|j)$, is  obtained by convolving the predicted distribution of intrinsic SZE signal with the SPT selection function given by the condition $\xi>4$,
\begin{equation}\label{eq:sptdetectionfraction}
f(\text{SPT}|j)=\frac{1}{\sum_f \Omega_f} \sum_f \Omega_f \int \text{d}\zeta\, P(\xi>4|\zeta) P(\zeta| j, \gamma_f), 
\end{equation}
where $\Omega_f$ is the solid angle of the field $f$ in the joint footprint. The weighted sum over the fields properly accounts for the 
spatially varying SPT-SZ survey depth.

Following from Eqns.~\ref{eq:detect} and \ref{eq:nondetect}  the effects of contamination is included by substituting 
\begin{equation}
f(\text{SPT}|j) \mapsto \big(1-\pi_\text{c}(\hat \lambda^j) \big) f(\text{SPT}|j),
\end{equation} 
where we choose $\hat\lambda^j$ as the geometrical mean of the $\hat\lambda_{+/-}^j$. 

\subsubsection{Mean mass in richness/redshift bin}
\label{sec:meanmass}

The prediction for the mean mass of clusters in an observed richness and redshift bin $j$ defined by $\hat\lambda_{-}^j<\hat\lambda<\hat\lambda_{+}^j$ and $z_{-}^j<z<z_{+}^j$ 
can be computed from the predicted distribution of masses
\begin{equation}
P(M|j) \propto  \int \text{d}z \int \text{d}\lambda\, P(j|\lambda, z) P(\lambda|M, z) \frac{\text{d} N}{\text{d}M}\Big|_{M,z}, 
\end{equation}
where the normalization is given by $\int \text{d}M\, P(M|j) = 1$. The mean mass $\bar M(j)$ is then simply 
\begin{equation}\label{eq:mean_mass}
\bar M(j) =  \int\text{d}M\, M P(M|j).
\end{equation}

Assuming that a fraction $\pi_\text{c}(\hat \lambda^j)$ of the objects in the bin are contaminants, we can investigate the mean mass of the contaminants $\bar M_\text{c}^j$. The mean measured weak lensing mass $\hat M_\text{WL}^j$  in this bin, as inferred in stacked WL analyses \citep{mcclintock19, des_y1_cluster}, can then be expressed as 
\begin{equation}\label{sec:meanwlmass}
\hat M_\text{WL}^j = \big(1-\pi_\text{c}(\hat \lambda^j)\big) \bar M(j)  +\pi_\text{c}(\hat \lambda^j) \bar M_\text{c}^j \pm \delta \hat M_\text{WL}^j,
\end{equation}
where $ \delta \hat M_\text{WL}^j$ is the reported measurement error on the mean mass.
We convert the measurements of the mean mass in redMaPPer richness and redshift bins within the DES collaboration \citep{mcclintock19,des_y1_cluster} to the mass definition of $M_{500c}$ by assuming an NFW mass profile and the mass concentration relation from \citet{bhattacharya13}. Together with the predictions on $\bar M(j)$ and  $\pi_\text{c}(\hat \lambda^j)$ from this work, we can predict the mean mass of the contaminants,
\begin{equation}\label{eq:meancontmass}
 \bar M_\text{c}^j = \bar M(j) + \pi_\text{c}^{-1}(\hat \lambda^j) \big( \hat M_\text{WL}^j  - \bar M(j) \big) \pm  \pi_\text{c}^{-1}(\hat \lambda^j)  \,\delta \hat M_\text{WL}^j.
\end{equation}
The difference between the contaminants mean mass $\bar M_\text{c}^j$ and the expected mean mass $\bar M(j)$ is thus sourced by the difference between the measured  WL mass $\hat M_\text{WL}^j$ and the expected mean mass $\bar M(j)$, modulated by the inverse of the contamination. Few contaminants with masses very different from the expected mass have the same impact on the measured mean mass as many contaminants  with masses closer to the mean expected mass.

 Physically, the hypothesis is that contaminants in optical cluster selection are low mass groups or even individual massive red galaxies residing in a line of sight crowded by other red galaxies at approximately the same photometric redshift.

\subsubsection{Number counts in richness/redshift bin}

The number of redMaPPer clusters $N(j)$ in a richness and redshift bin $j$ defined by $\hat\lambda_{-}^j<\hat\lambda<\hat\lambda_{+}^j$ and $z_{-}^j<z<z_{+}^j$ is 
given by 
\begin{eqnarray}\label{eq:numberofRMclusters}
N(j) = & \int \text{d}z\,  \Omega(z)\,\int \text{d}\lambda\, P(j|\lambda, z)  \\
& \int \text{d}M\, P(\lambda|M, z) \frac{\text{d} N}{\text{d}M}\Big|_{M,z}, 
\end{eqnarray}
where $\Omega(z)$ is the solid angle in units of degrees in which an object with redshift $z$ can be detected \citep{des_y1_cluster}. This is expressed as a function of redshift to encode that redMaPPer clusters are only presented within particular redshift ranges.  
We compare 
our prediction for the number counts to other code available within DES \citep{des_y1_cluster} and find that the numerical differences are clearly smaller than the 
systematic uncertainties on the quantity derived from marginalizing over our richness--mass relation and the cosmological parameters.

In the case of contamination, we can assume that the total number of objects $N_\text{tot}(j)$ in the bin $j$ is the sum of the number of clusters $N(j)$ and contaminants $N_\text{c}(j)$. Using 
$N_\text{c}(j) = \pi_\text{c}(\hat \lambda^j) N_\text{tot}(j)$, we find that
\begin{equation}\label{eq:numberofRMclusters_cont}
N_\text{tot}(j) = \frac{N(j)}{1- \pi_\text{c}(\hat \lambda^j)} = \big(1+A(\hat\lambda^j)\big) N(j),
\end{equation}
where the last transformation is made using the parametrisation of the richness dependence of the contamination given by Eq.~\ref{eq:picont_lambda}. This provides a physical interpretation for that parametrisation: $A(\hat\lambda)$ is the ratio between the number of contaminants and the number of clusters at the richness $\hat \lambda$, as $N_\text{c}(j) = A(\hat\lambda^j) N(j)$.


\begin{figure*}
	\includegraphics[width=\textwidth]{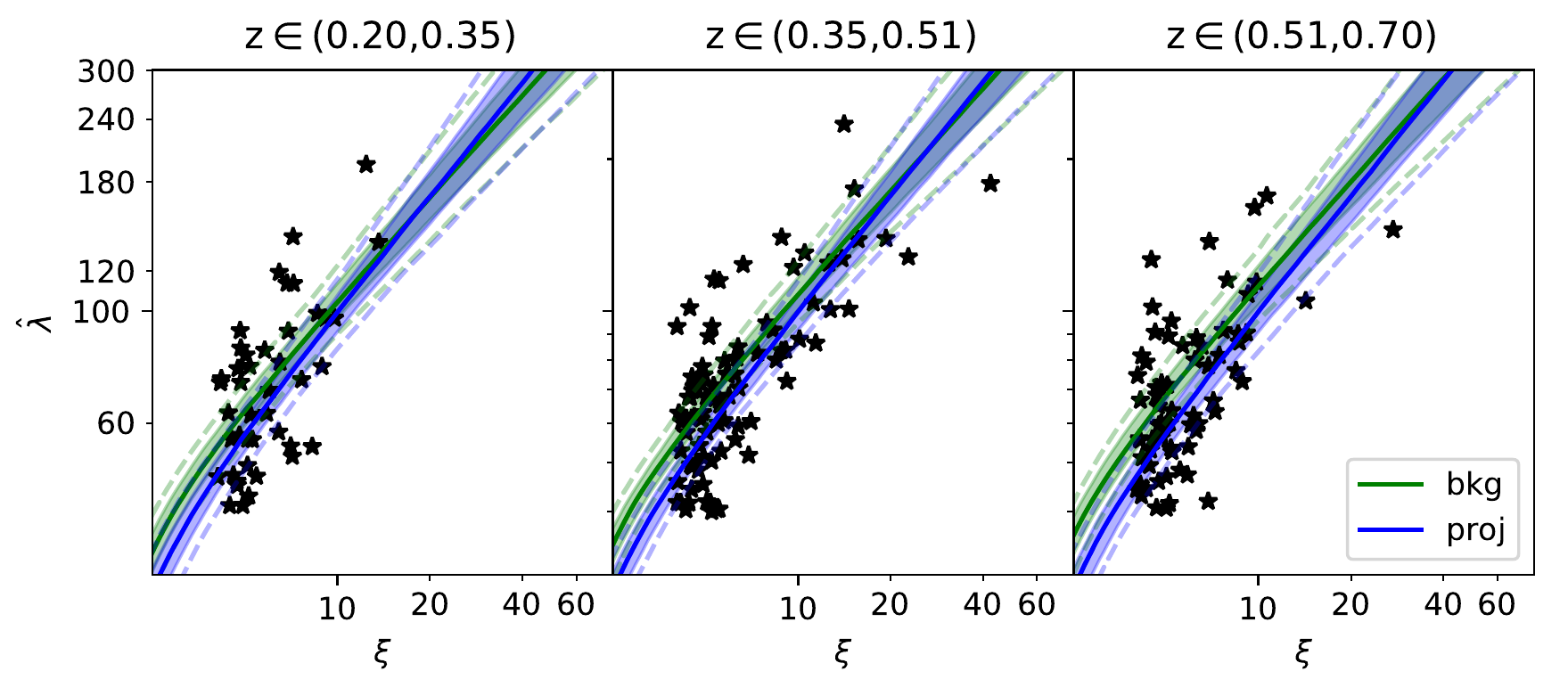}
	\vskip-0.10in
    \caption{The richness as a function of SZE signal of the matched sample (black dots) in three different redshift bins with mean redshift increasing from left-hand panel to right-hand panel. 
    Over-plotted is the mean richness-SZE signal relation and associated 68\% (filled area), and 95\% (dashed line) uncertainties resulting from the analyses assuming the two error models (blue: projection model, green: background only model). The small deviation of solid-blue lines from solid-green lines reflect the impact of projection effects of the mean relations.}
    \label{fig:showing_data}
\end{figure*}

\subsection{Code and model validation}\label{sec:validation}

We validate our code in two stages. First, we test for potential biases induced by the model fitting. To do this, we create synthetic datasets drawn from our models assuming specific input parameters.  We then run our likelihood analysis on the synthetic data and  compare the posteriors on the parameters to the known input values. Here we draw a synthetic cluster sample in a footprint that is 10 times larger than the actual survey footprint. This reduces statistical noise by a factor of $\sqrt{10}$. With this approach we demonstrate that the code recovers the input parameters at better than 1 $\sigma$. Biases in our analysis of the real data are thus smaller that $\sim0.3\sigma$.

The second stage of the validation tests whether the adopted model is an adequate description of the observations. This question in an inherently scientific question that can only be answered by analysing the actual data. In the course of this work we then pursue three lines of argument to assess the adequacy of our models. First, we investigate if our model provides a good fit to the data. Given that our likelihoods are not Gaussian linear models, we can not use a $\chi^2$-test. Generalizations of the $\chi^2$-test to arbitrary Bayesian likelihood analysis are beyond the scope of this work, so here we employ visual comparison between the data and the model prediction. In order to generalized  $\chi^2$-test to arbitrary Bayesian likelihood analysis, one needs to draw a large amount of mock data sets, and run full likelihood analyses on them \citep{nicola19, doux21}. Our prior experience with simplified versions of such methods in \citet{bocquet15, bocquet19} revealed that in all cases where advanced Bayesian method detected "bad" fits to the data, this was also abundantly clear by visual inspection. Furthermore, these methods are of questionable efficacy from a statistical and philosophical prospective \citep{kerscher19}. In the cases of interest to this work, specifically the question whether our prediction for the SPT confirmation fraction matches the measured confirmation fraction, visual inspection provides sufficient discriminatory power. Another element to assess the adequacy of the model is comparison to independent external results. Lastly, if a model is physically plausible, this provides further support for it, or, conversely, if the model makes implausible predictions, that would provide reason to reject it. Either way, exploring which predictions your model makes, and how these could be tested, is of importance for its validation.

\section{Results}

In this section we present our main results, starting with the calibration of the richness--mass relation based on the SPT 
cross matching. We then proceed to constrain the redMaPPer contamination fraction. We finally seek to extend the 
measurement of the redMaPPer contamination to lower richnesses by comparing measured and predicted number counts of optically-selected clusters and their stacked WL signal. 

\begin{figure*}
	\includegraphics[width=\textwidth]{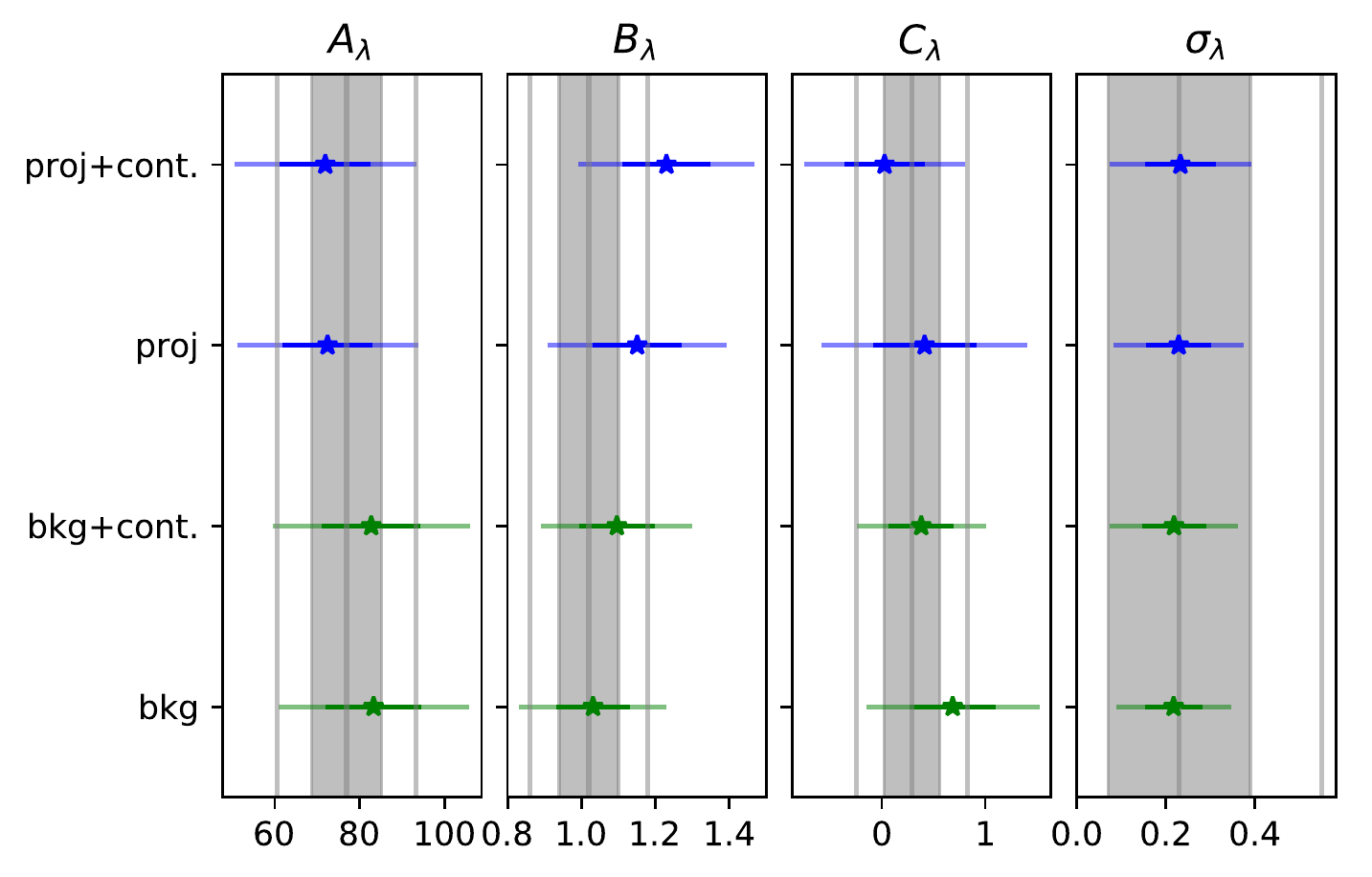}
	\vskip-0.10in
    \caption{Means and standard deviations estimated from the single parameter posteriors for the 
	richness--mass scaling relation when cross-calibrated using priors on the SZE--mass relation
	under the assumption of two different $\hat\lambda$ measurement noise models: (1) `background' in green
	(Eq.~\ref{eq:errormodellambda}) and (2) `projection' in blue \citep{costanzi19}.  
	`+ cont': includes confirmation fraction (richness-dependent failure fraction of the richness-mass scatter model). The full errorbars denote one standard deviation, while the shaded ones denote 2 standard deviations
	Gray bands denote the mean and standard deviation
	of the richness--mass relation parameters reported by \citet{bleem19}, with the vertical lines denoting two standard deviations. The results of the
	different analyses are mutually consistent.}
    \label{fig:proj_both_probs}
\end{figure*}

\begin{table*}
	\centering
	\caption{Mean and standard deviation estimated from the single parameter posteriors for the 
	richness--mass scaling relation when cross-calibrated using priors on the SZE--mass relation
	under the assumption of two different $\lambda$ measurement noise models: (1) background
	(Eq.~\ref{eq:errormodellambda}) and (2) projection \citep{costanzi19}.
	`+ cont': includes confirmation fraction. 
	`--' appears for the optical--SZE scatter correlation coefficient $\rho$, because the parameter is 
	not constrained. }
	\label{tab:results}
	\begin{tabular}{lccccccc} 
		\hline
		 & $A_\lambda$ & $B_\lambda$& $C_\lambda$ & $\sigma_\lambda$ & $\rho$  &  $A_\text{c}$ & $B_\text{c}$\\
		\hline
		background & &  & &  &  &  &\\
		\hline
		SPT calibr & 83.3$\pm$11.2 & 1.03$\pm$0.10 & 0.69$\pm$0.42 & 0.22$\pm$0.06 & -- &   & \\
		+ cont & 82.7$\pm$11.6 & 1.06$\pm$0.10 & 0.38$\pm$0.31 & 0.22$\pm$0.07 & -- & 0.94$\pm$0.58 & -1.69$\pm$0.89\\
        \hline
		projection & &  & &  &  &  &\\
		\hline
        SPT calibr & 72.4$\pm$10.6 & 1.15$\pm$0.12 & 0.41$\pm$0.50 & 0.23$\pm$0.06 & -- &  & \\
		+ cont & 71.9$\pm$10.7 & 1.23$\pm$0.12 & 0.03$\pm$0.39 & 0.23$\pm$0.08 & -- &  1.15$\pm$0.66 & -2.26$\pm$1.01\\
	\end{tabular}
\end{table*}

\subsection{Cross calibration of $\lambda$-mass relation}\label{sec:crosscalibration}

We present our calibration of the richness--mass scaling relation parameters and then examine what our sample can tell us about the sample contamination.

\subsubsection{Scaling relation parameters}

As outlined in Section~\ref{sec:SPTcc} the measured richnesses and SZE signals for clusters in the cross matched redMaPPer-SPT 
sample enable us to transfer the calibration of the SZE signal--mass relation 
given by published priors \citep{bocquet19} to the richness--mass relation. 
The scatter plot of the matched sample in 
redshift bins is shown in Fig.~\ref{fig:showing_data} with black points. 

The resulting posteriors on the parameters of the richness--mass scaling relation are summarized in Table~\ref{tab:results} and in 
Fig.~\ref{fig:proj_both_probs}. In blue we show the constraints from adopting the projection optical error model and sampling the cross 
calibration likelihood, while in green we show the posteriors when using the background optical error model. The constraints are in very good agreement, highlighting that at the level of statistical constraining 
power of the cross matched sample the two error models are not distinguishable. The minor changes induced by the projection 
effects can be seen by comparing the two mean relations shown in Fig.~\ref{fig:showing_data}. For comparison in Fig.~\ref{fig:proj_both_probs} we also plot in grey bands previous 
results by \citet{bleem19} from analysing the redMaPPer richness--mass relation of the SPT-SZ and the SPTpol Extended Cluster Survey. This analysis is not independent on our work as the data partially overlap: \citet{bleem19} used the entire SPT-SZ sample together with the SPTpol Extended Cluster Survey and DES Y3 data, while we use DES Y1 data and restrict ourselves to the part of the SPT-SZ sample in the DES-Y1 footprint. The good agreement with our results is thus more a consistency check for the accuracy of our analysis method than an assessment on the adequacy of the assumed model.

\begin{figure}
	\includegraphics[width=\columnwidth]{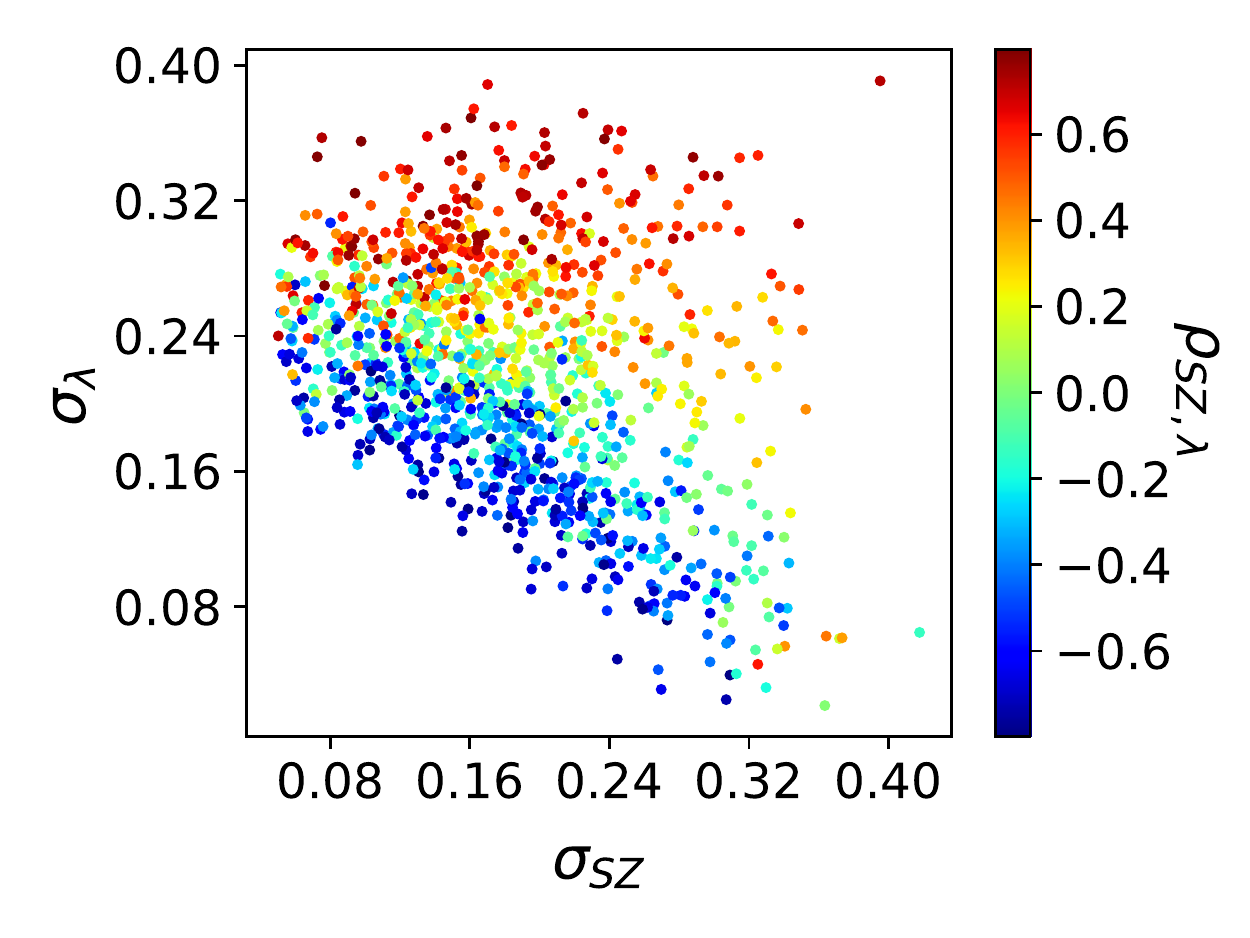}
	\vskip-0.10in
    \caption{The data primarily constrains the total richness-SZE signal scatter, so there are significant degeneracies between the three scatter terms in the model, the intrinsic scatters in richness and SZE signal as well as the correlation coefficient between these two scatters. Assuming a prior on the SZE scatter, constraints on the correlation coefficient are not to be expected. We illustrate this degeneracy by plotting a random subset of the posterior chain of the background only cross calibration analysis.}
    \label{fig:scatter_corrcoeff}
\end{figure}

\subsubsection{Correlated scatter}\label{sec:correlated_scatter}

The correlation coefficient $\rho$ between the scatters in observables carries important astrophysical information on the physical processes inside galaxy clusters \citep[as shown in simulations by ][]{stanek10, angulo12, wu15, farahi18, truong18}. Ignoring this correlation in cluster population studies can lead to biases in the inferred scaling relation \citep{allen11, angulo12}. In this work, the correlation coefficient is free to vary between -1 and 1. Unfortunately, our data do not provide any significant constraint on $\rho$, because our data only determine the total scatter among two observables. While the observational contribution to that total scatter can be accounted for, the total intrinsic scatter in richness at a given SZE signal is
\begin{equation}
\sigma^2_\text{tot} \approx \sigma^2_\lambda + \Big( \frac{B_\lambda}{B_\text{SZE}} \sigma_\text{SZE} \Big)^2 - 2 \rho \sigma_\lambda \Big( \frac{B_\lambda}{B_\text{SZE}} \sigma_\text{SZE} \Big).
\end{equation}
Imposing a prior on the intrinsic scatter $\sigma_\text{SZE}$ and the mass trend $B_\text{SZE}$ of the SZE observable, one can constrain the slope $B_\lambda$ of the richness--mass relation from the average trend in the data (cf. Fig~\ref{fig:showing_data}). This results in a degeneracy between the intrinsic scatter in richness $\sigma_\lambda$ and the correlation coefficient $\rho$, which we show in Fig.~\ref{fig:scatter_corrcoeff}. Larger intrinsic scatter in richness can be accommodated  by positive correlation coefficients, while smaller scatter requires negative correlation coefficients. Due to this fundamental degeneracy, we do not expect to measure the correlation coefficient. Recent reported detections of correlations in the scatter of two observables \citep{mulroy19, farahi19b} are likely sourced by the assumption that the intrinsic scatter of the weak lensing inferred mass is perfectly known. The lack of a detection of the correlation coefficient in this work thus reflects a more refined handling of systematic uncertainties.

\begin{figure*}
	\includegraphics[width=\textwidth]{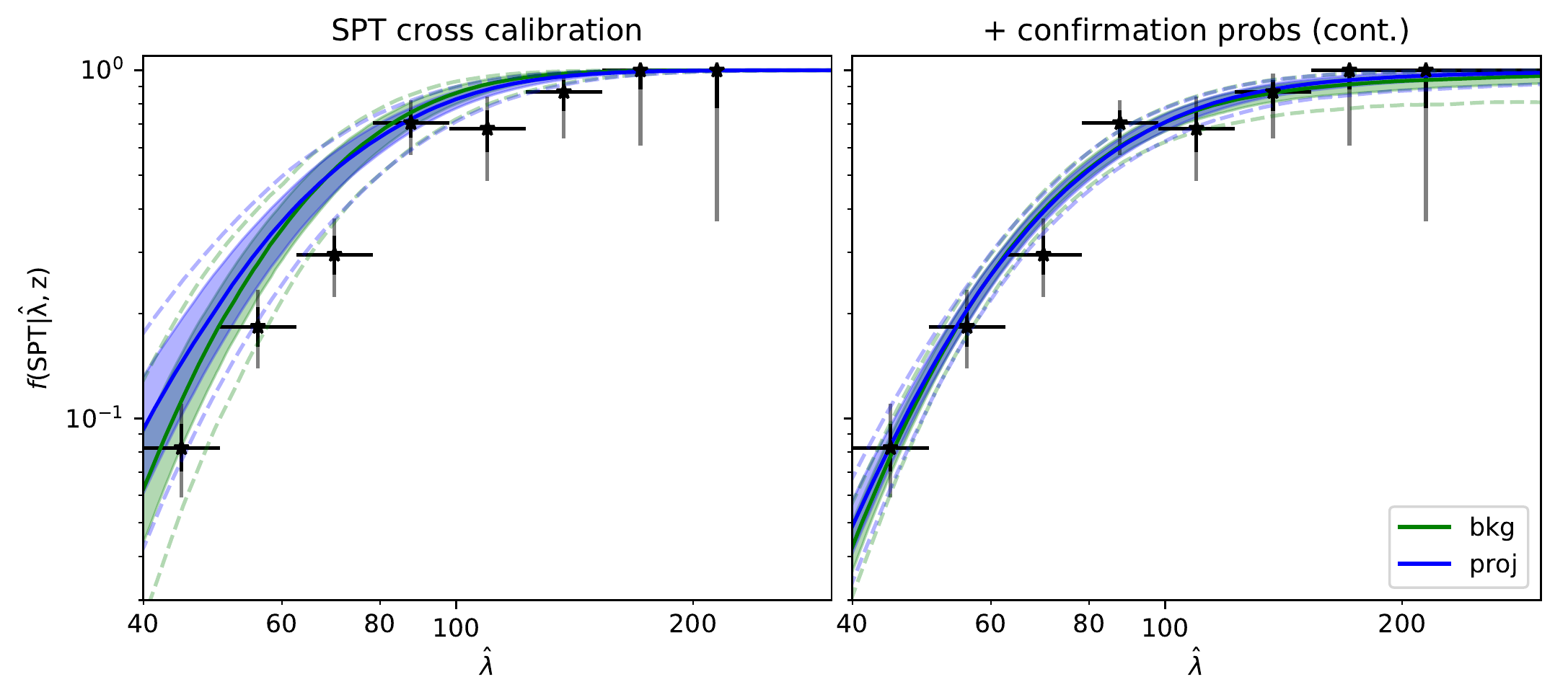}
	\vskip-0.10in
    \caption{SPT confirmation fraction as a function of measured richness and associated 68\% (filled area), and 95\% (dashed line) uncertainties, predicted from the 
    posteriors on the scaling relation parameters for the background error model (`bkg', green) and when considering projection effects 
    (`proj', blue), overlaid with the measurement as black points (black error bars are 1 $\sigma$, grey error bars 2 $\sigma$ uncertainty). Left-hand panel shows the predictions for the SPT calibration only, and the right-hand panel shows the addition of the confirmation fraction likelihood in the contamination case. Clearly, the confirmation fraction resulting from our richness--mass model alone is insufficient to predict the actual confirmation fraction. We interpret this as the presence of a population of redMaPPer selected objects for which our scatter model fails. We call this population contaminants. The contamination fraction is richness dependent and provides a better fit to the SPT confirmation fraction of redMaPPer objects, while the choice of 'bkg' or 'proj' makes little difference. }
    \label{fig:SPT_conf_fraction}
\end{figure*}

\begin{figure}
	\includegraphics[width=\columnwidth]{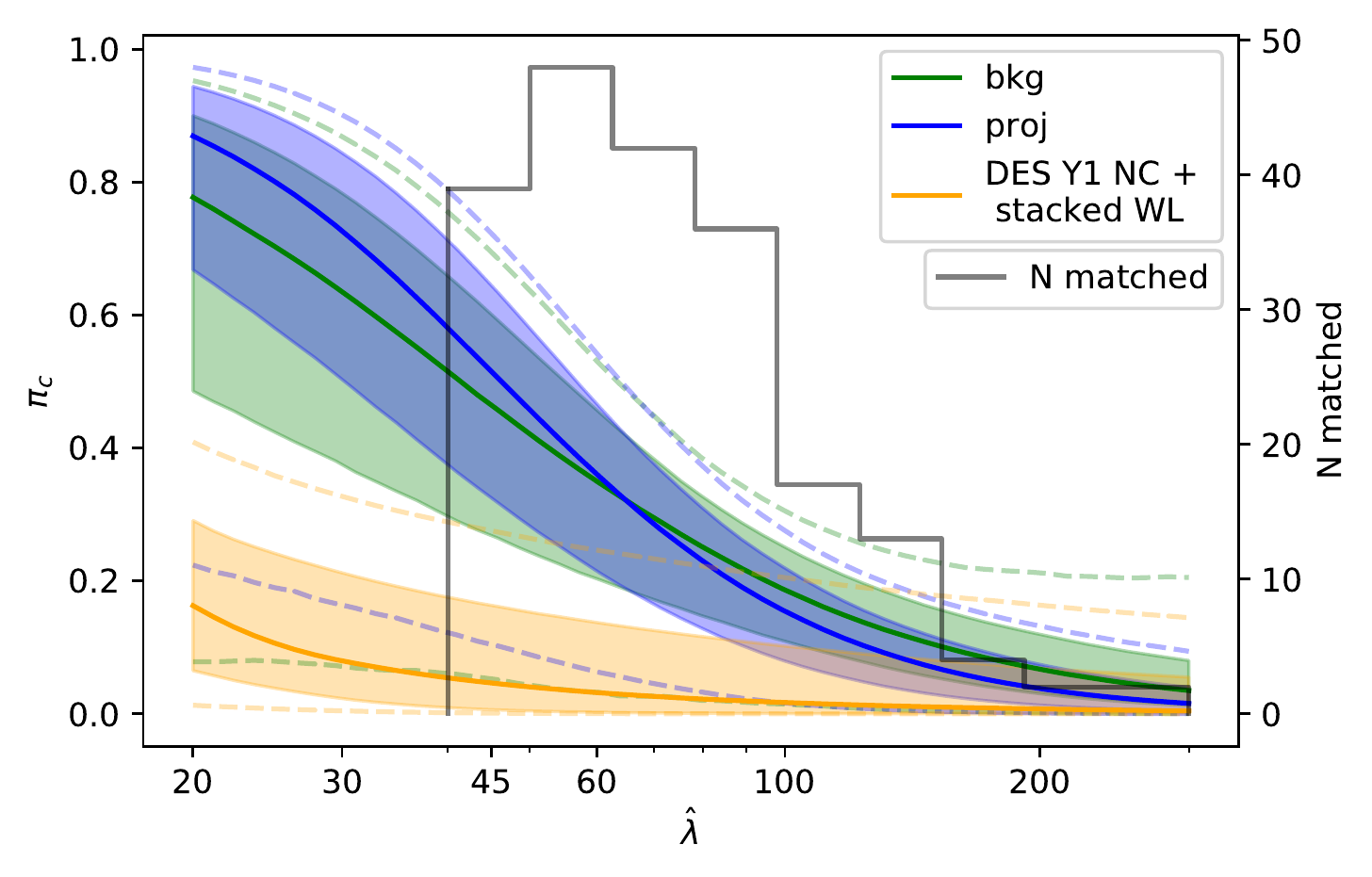}
	\vskip-0.10in
    \caption{Constraints on the fraction of failures in our richness--mass scatter model, which we call contaminants, as a function of measured richness from our analyses of the SPT confirmation fraction of redMaPPer objects with two different optical error models (blue: accounting for projection effects, green: background only), with associated 68\% (filled area), and 95\% (dashed line) uncertainties. In yellow the result from \citet{des_y1_cluster}, when fitting mean weak lensing masses and number counts of redMaPPer objects with cosmological priors from the shear and red galaxy auto- and cross-correlations \citep{desY1_3x2pt}. These measures are consistent with our results at less than 2 $\sigma$. In grey, the richness distribution of redMaPPer systems confirmed by SPT, indicating that the bulk of our constraining power comes richnesses $\hat\lambda\sim60$. }
    \label{fig:cont_of_lambda}
\end{figure}

\subsubsection{SPT confirmation probabilities}

As discussed  in Section~\ref{sec:outliers_and_purity} the probability of confirming a redMaPPer 
cluster in SPT is very sensitive to the respective scaling relation parameters of the two selection
observables. Consequently, one needs to 
marginalize over a reasonable range of scaling relation parameters when inferring the 
contamination level of the redMaPPer sample with the likelihood given in Eq.~\ref{eq:probs_like} (see also Eq.~\ref{eq:confirmationprobability}). To do this, we sample the likelihood of 
confirmation probabilities simultaneously with the cross calibration likelihood (Eq.~\ref{eq:crosscalib_lnL}). This ensures proper accounting for the
systematic uncertainties on the scaling relations when inferring the contamination fractions. The resulting posterior, 
depending on the assumed optical error model, are shown in green and blue in Fig.~\ref{fig:proj_both_probs} for the  
background (`bkg'; Eq.~\ref{eq:errormodellambda}) and projection \citep[`proj';][]{costanzi19} model, respectively, and summarized in Table~\ref{tab:results}. 

The posteriors on the scaling relation parameters are generally in good agreement with the 
results without the confirmation 
probability likelihood.
The detection probability likelihood slightly alters the posteriors on the 
scaling relation parameters, having the largest impact on the mass trend and the redshift trend. 
This is not surprising when considering that not detecting a redMaPPer 
object in SPT is equivalent to the measurement $\xi<4$, which given priors on the SZE signal--mass relation carries some mass 
information, at least in the form of an upper limit. This information is however quite weak, 
as can be seen by examining the change in measurement uncertainties. Furthermore, it is 
consistent with the information recovered from the 
matched sample alone, because the shifts is mean values also do not exceed 1 $\sigma$.

Using Eq.~\ref{eq:sptdetectionfraction}, we predict the SPT confirmation fraction of the redMaPPer ($\hat\lambda>40$) sample as a 
function of measured richness, shown in Fig.~\ref{fig:SPT_conf_fraction}. The shaded region (dashed line) reflects the 1 (2) $\sigma$ systematic uncertainty as 
propagated from the posterior samples for the background error model (green), and the projection model (blue).  We also plot the measured confirmation rate as black point with errorbars. Note also here that the difference between the predictions for the two error models is
small. In the left panel, we show the prediction for the baseline SPT calibration with no contamination, while in the right panel we show the case of richness dependent contamination The case without contamination tends to over-predict the confirmation fraction. In contrast, the richness dependent contamination provides a better description of the data. Thus, the case with richness dependent contamination provides a better fit to the SPT confirmation of redMaPPer objects with measured richness $\hat \lambda>40$, than the case without contamination.

In Fig.~\ref{fig:cont_of_lambda} we show our posterior on the contamination fraction as a function of richness for the two optical error models considered (blue: accounting for projection effects, green: background only). The two posteriors are in good agreement with each other when considering the uncertainties we derive. We also show in grey the number of SPT-SZ confirmed redMaPPer. This indicates that the bulk of the constraining power of the confirmation fraction comes from richnesses $\hat\lambda\sim60$. We also present the indirect constraint on the contamination fraction that has been derived independently \citep{des_y1_cluster} using the number counts of redMaPPer objects, mean weak lensing mass measurements of redMaPPer objects, and cosmological priors from the shear and red galaxy auto- and cross-correlations \citep{desY1_3x2pt}. That analysis predicts a significantly lower contamination than our results at about the 2$\sigma$ level. 

Importantly, the external analysis \citep{des_y1_cluster} assumes that the mean contaminants have zero mass, and the contamination fraction they derive provides a poor fit to the mean WL masses in the lowest richness bins of the redMaPPer sample. Furthermore, assuming the lower contamination fraction as priors, the cosmological constraints from refitting the number counts and mean masses were not in agreement with the cosmological constraints from the shear and red galaxy auto- and cross-correlations \citep{desY1_3x2pt}. In summary, there are apparent qualitative and quantitative differences between our constraints on the redMaPPer contamination fraction, and those from the external analysis \citep{des_y1_cluster}. 
The contamination levels we find at lower richness
are surprisingly large. We will demonstrate in the following that these high contamination levels are physically plausible.

Finally, we stress that `contaminants' in this work simply reflect objects that at a given richness do not conform with our SZE derived mass distribution. The Bayesian population model allows us to perform such a derivation while taking account of several biases (mainly the \textit{Eddington bias} \citep[e.g.][]{mortonson11} and the  \textit{Malmquist bias}  \citep[e.g.][]{allen11}). Both of these biases depend on scatter between observables and mass that occures also below the selection thresholds. Extrapolating an inadequate scatter model can thus bias the mean relation and alter our definition of `contaminant'. An instance of such an effect is described in the discussion section below (cf. Section~\ref{sec:alternatives}).

\subsection{Comparison to redMaPPer observables}\label{sec:extrapolation}

In the previous section we determined the systematic uncertainty on the richness--mass relation in the regime of measured 
richness >40, which is at the intermediate to high mass end. We also established that a sizeable amount of contamination provides a better fit to the observed SPT confirmation fraction than alternative models. In this section we address whether such large contamination fractions are consistent with the measured number of redMaPPer objects as a function of richness and redshift. We then employ the mean mass from stacked weak lensing measurements in richness--redshift bins to estimate the mean mass of the contaminants. We consider richness and redshift bins with edges at $\lambda\in (20,30,45, 60,200)$ and redshift bins with edges 
$z\in(0.20,0.35,0.50,0.60)$  \citep[the same used in][]{des_y1_cluster}, from which we also take the measured masses and uncertainties.

\begin{figure}
	\includegraphics[width=\columnwidth]{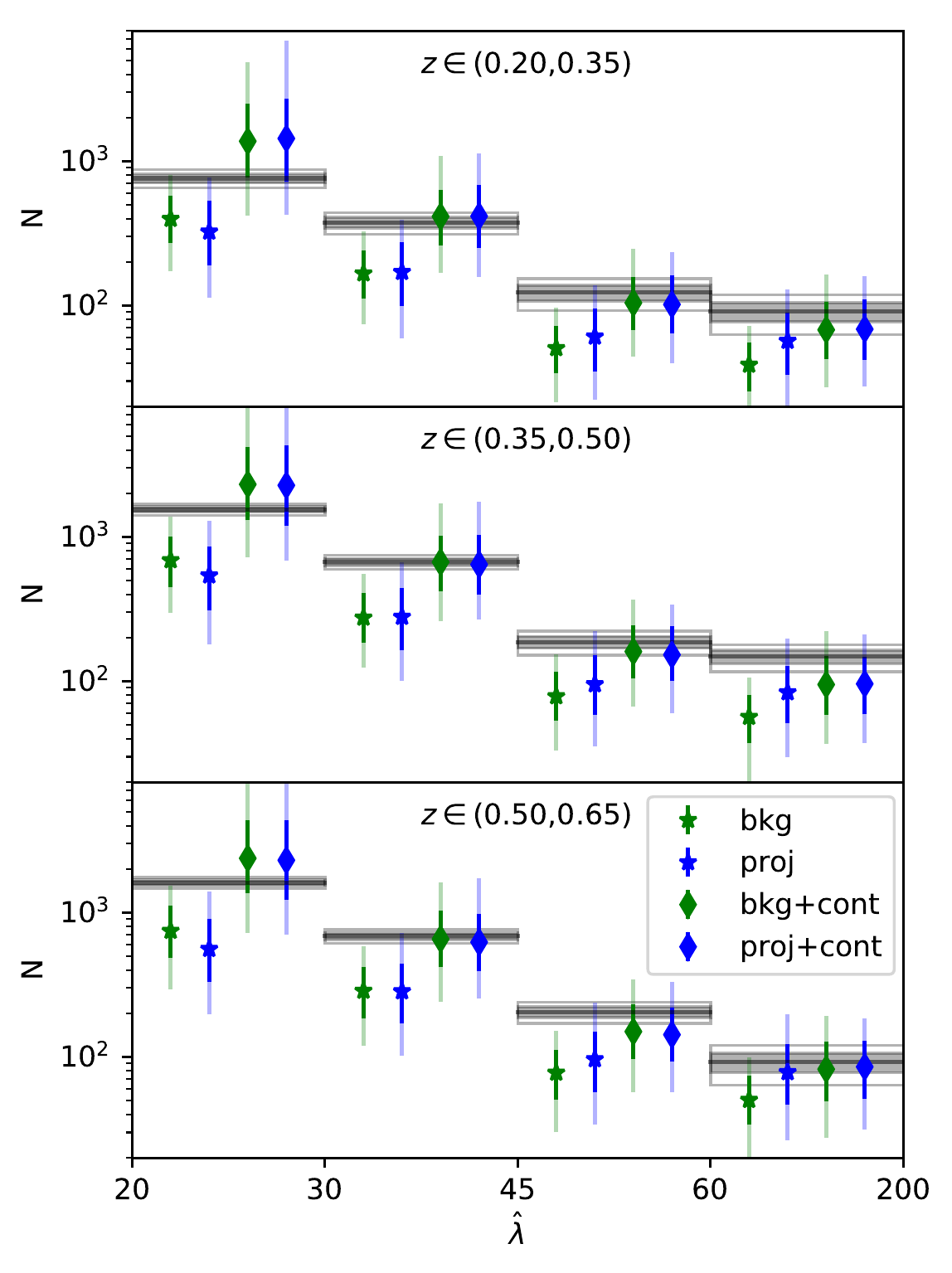}
	\vskip-0.10in
    \caption{The number of redMaPPer objects in bins of richness and redshift are shown as predictions (colored points) derived from our richness--mass relation 
    constraints and SPT cluster number counts cosmology for the different optical error models (blue projection, green background). Full error bars show the 16th/84th-percentile, while shaded error bars show the 2.5th/97.5th percentile. Stars denote predictions without contamination, while diamonds denote predictions with contamination. Over plotted in grey is the number of redMaPPer objects with the associated statistical uncertainty (empty boxes showing the 2$\sigma$ uncertainty).  When considering the systematic error bars, the prediction with richness dependent contamination (see Fig.~\ref{fig:cont_of_lambda}) provides a better match to the data than the prediction without contamination.
    }
    \label{fig:comp_NC_marg}
\end{figure}

\subsubsection{Number counts}\label{sec:RM_numbercounts}

We predict the number of redMaPPer clusters in richness and 
redshift bins using the posteriors on the richness-mass relation derived with and without the contamination fraction model (equations~\ref{eq:numberofRMclusters}~and~\ref{eq:numberofRMclusters_cont}, respectively). As stated 
above, we not only propagate the uncertainty on the richness--mass scaling relation 
parameters, but also those on the 
cosmological parameters by sampling the cosmological parameters within the priors reported in Section~\ref{sec:SPTcc}. This results in a prediction of the number of objects with large systematic uncertainties, as shown in Fig.~\ref{fig:comp_NC_marg} for the projection model (blue), and the background model (green). Stars denote the prediction assuming no contamination, while diamonds denote the prediction accounting for our constraints on the contamination. In the same figure we also plot as gray bands the number of redMaPPer objects with their statistical uncertainties \citep{des_y1_cluster}, which are considerably smaller than the systematic uncertainties. 

The predictions with richness dependent contamination provide a better description than the predictions without contamination. This shows that the contamination fraction that we fit is not in contradiction with the measured number of redMaPPer clusters. While uncertainties on the contamination fraction are still large, especially the lower limit we place on the contamination fraction is clearly compatible with measured number counts. Conversely, the maximum likelihood value of our contamination posterior over-predicts the number counts in the lowest richness bin. This statement also holds true if instead of the SPT cosmology \citep{bocquet19}, we use the cosmological constraints from the study of the cosmic shear and galaxy auto- and cross-correlations in DES-Y1 data \citep{desY1_3x2pt}, as shown in Appendix~\ref{app:3x2pt}. We would like to stress here, that we do not fit for the number counts or redMaPPer objects, but that we extrapolate our scaling relation (that is calibrated for $\hat\lambda>40$) and SPT cosmological results (derived for $M \gtrsim 3 \times 10^{14} \text{M}_\odot$) to lower masses and richnesses. Indeed, the better agreement of external data with our prediction in the presence of contamination provides further support for the presence of a considerable richness dependent contamination in the redMaPPer sample. 

\begin{figure}
	\includegraphics[width=\columnwidth]{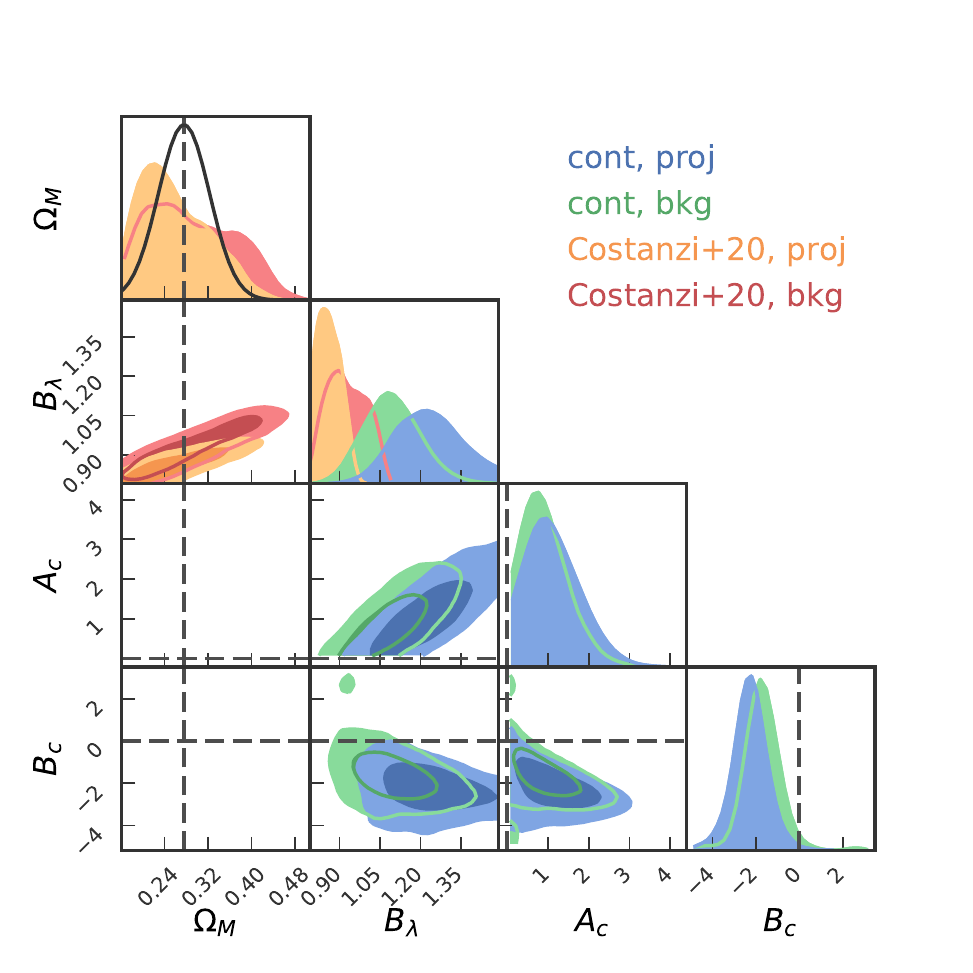}
	\vskip-0.10in
    \caption{Marginal posteriors on the matter density $\Omega_\text{M}$, the mass trend of the richness mass relation $B_\lambda$ and the parameters amplitude $A_\text{c}$ and richness trend $B_\text{c}$ of the contamination fraction from our work (green: 'bkg', blue: 'proj'), in comparison with the joint analysis of the SPT multi-wavelength data and the redMaPPer number count by \citet{costanziip} (red and yellow), which did not fit the SPT confirmation fraction. Black is the $\Omega_\text{M}$ posterior from \citet{bocquet19}. Clearly, the amplitude of the contamination fraction is degenerate with the mass trend of the richness. In the context of redMaPPer number counts fits, the matter density and the richness slope are degenerate.}
    \label{fig:degens}
\end{figure}

An interesting comparison can be drawn with the work by \citet{costanziip}. In that work, the redMaPPer number counts were fitted jointly to the SPT WL follow-up data \citep{bocquet19} that was used to derive the priors on the SZ scaling relation parameters for our analysis, while the SPT confirmation fraction was not fit. A strong degeneracy between the matter density $\Omega_\text{M}$ and the mass trend of the richness, $B_\lambda$, was found, as shown in Fig.~\ref{fig:degens}. Note that in that work, the authors did not entertain the possibility that the richness--mass model would fail, leading to contamination. When comparing to their constraints on the richness-mass trends, we find them to be lower than but consistent with our results. Assuming the same matter density prior as we did (Fig.~\ref{fig:degens}, in black) would improve the agreement. More interestingly, we find in our analysis of the SPT confirmation fraction of redMaPPer objects a strong degeneracy between the amplitude of contamination $A_\text{c}$ and the mass trend of the richness $B_\lambda$, suggesting that shallower mass trends would lead to less contamination. Furthermore, a degeneracy between the contamination fraction and the cosmological constraints is to be expected when fitting the number counts. Future joint analysis of the SPT confirmation fraction and the redMaPPer number counts are expected to correctly explore these degeneracies in the likelihoods while plausibly putting tighter constraints on the contamination fraction.

\begin{figure}
	\includegraphics[width=\columnwidth]{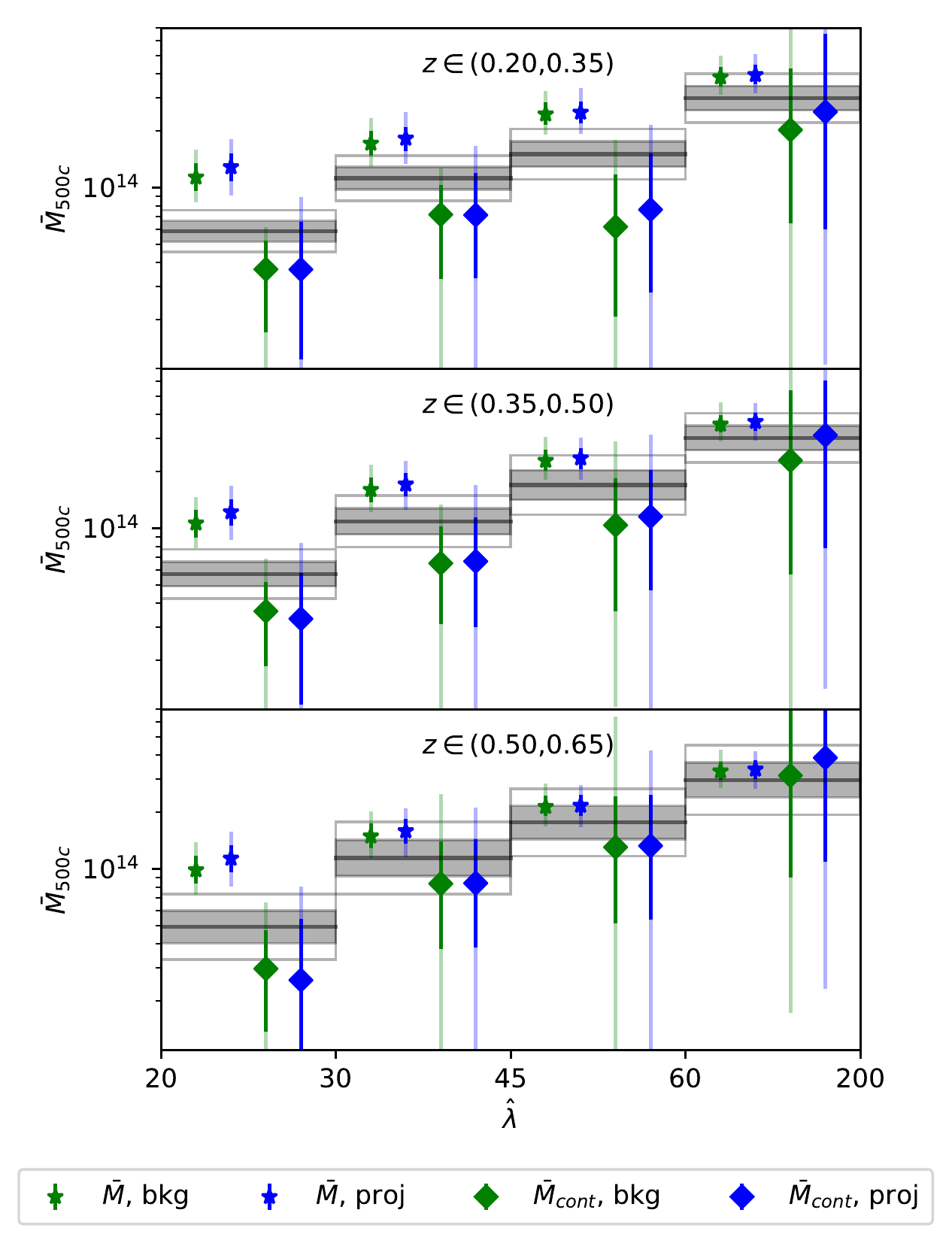}
	\vskip-0.10in
    \caption{Mean mass of the clusters (stars) and contaminants (diamonds) in redshift and richness bins predicted from our posteriors for the background error model (green), and the projection 
    error model (blue). Full error bars show the 16th/84th-percentile, while shaded error bars show the 2.5th/97.5th percentile. Over plotted as gray bands  the mean 
    masses reported by \citet{des_y1_cluster} from stacked weak lensing (empty boxes showing the 2$\sigma$ uncertainty).}
    \label{fig:comp_McClintock_marg}
\end{figure}

\begin{table}
	\centering
	\caption{Mean cluster masses $\bar M$, contamination fractions  $\pi_\text{c}(\hat\lambda^j)$ and resulting mean contaminants masses  $\bar M_\text{c}$ in richness--redshift bins. These values reflect the systematic uncertainty on the richness-mass relation and the richness dependent contamination from the posteriors of the SPT calibration and the SPT confirmation probability fits for the background only model.}
	\label{tab:meancontm}
	\begin{tabular}{l|cccc} 
		\hline
		richness bins & $(20,30)$ & $(30, 45)$& $(45, 60)$ & $(60, 200)$ \\
		\hline
		redshift bins & & &  &  \\
		\hline
	    $\log_{10} \bar M$ &  &  &  &  \\
	    $(0.2,0.35)$ & 14.06$\pm$0.07 & 14.23$\pm$0.06 & 14.39$\pm$0.06 & 14.58$\pm$0.05 \\
	    $0.35,0.50)$ & 14.02$\pm$0.07 & 14.20$\pm$0.06 & 14.36$\pm$0.06 & 14.55$\pm$0.05 \\
	    $(0.50,0.60)$ & 14.00$\pm$0.07 & 14.17$\pm$0.06 & 14.33$\pm$0.06 & 14.52$\pm$0.05 \\
	    \hline
	    $\pi_\text{c}(\hat\lambda^j)$ &  0.66$\pm$0.21& 0.51$\pm$0.18 & 0.38$\pm$0.14 & 0.16$\pm$0.06  \\
	    \hline
	    $\log_{10} \bar M_\text{c}$ &  &  &  &  \\
	    $(0.2,0.35)$ & 13.5$\pm$0.3 & 13.7$\pm$0.3 & 13.7$\pm$0.5 & 14.1$\pm$0.6 \\
	    $(0.35,0.50)$ & 13.5$\pm$0.3 & 13.7$\pm$0.4 & 13.9$\pm$0.4 & 14.3$\pm$0.5 \\
	    $(0.50,0.60)$ & 13.5$\pm$0.3 & 13.8$\pm$0.3 & 14.0$\pm$0.4 & 14.4$\pm$0.6 \\
	\end{tabular}
\end{table}

\subsubsection{Estimating the mean mass of contaminants}\label{sec:cont_masses}

As discussed earlier, our richness dependent contamination model provides a good fit to the SPT confirmation fraction of redMaPPer objects, as well as an improved prediction for the number of redMaPPer objects in richness and redshift bins. Knowing the mean mass of the clusters from our SPT calibration and the contamination fraction from our study of SPT confirmations we can employ the measured mean mass in richness--redshift bins to estimate the mean mass of the contaminants (Eq.~\ref{eq:meancontmass}). Mean cluster masses, contamination fractions and resulting mean contaminant masses are reported in Table~\ref{tab:meancontm} and in Fig.~\ref{fig:comp_McClintock_marg}. In the lowest richness  bin $20<\hat\lambda<30$, we find that 66\%$\pm$21\% of the objects are contaminants with a mean mass of $\sim3\times10^{13} M_\odot$, compared to a cluster population with a mean mass of $\sim1\times10^{14}M_\odot$.  These constraints vary little with redshift. In the next lowest bin ($30<\hat\lambda<45$), the contaminants make up 51\%$\pm$18\% of the population and have a mean mass of $\sim 5\times10^{13}M_\odot$, compared to the cluster populations mean mass of $\sim 1.6\times10^{14} M_\odot$. These mean masses of clusters and contaminants are roughly located at the extremes of the mass constraints derived from stacked weak lensing \citep[][see Fig.~9]{des_y1_cluster}, supporting their overall physical plausibility.  

For the higher richness bins the central value of the mean contaminants mass approaches the expected mean mass. In light of equation Eq.~\ref{eq:meancontmass}, this is the natural consequence of the smaller difference between the expected mean mass and the measured mean WL mass. Furthermore, the errorbars on the mean contaminants mass become very large at higher richness. As can be seen again in Eq.~\ref{eq:meancontmass}, the error scales like the inverse contamination fraction $\pi_\text{c}(j)^{-1}$. At larger richness, the contamination fraction tends to zero, increasing the error on the mean contaminants mass. 

From a physical perspective, it is both convenient and reasonable to refer to these objects as "contaminants", because their masses indicate that these are galaxy groups rather than galaxy clusters. Our measurement is consistent with the notion that every optically-selected object is associated with a halo of some mass. We find that for the redMaPPer sample, these contaminants constitute a significant fraction of the objects and have masses lying in the group mass range ($\sim3$-$5\times 10^{13} $ M$_\odot$).

\section{Discussion}\label{sec:discussion}

In the following we discuss the major results of our work. We first compare our results on the scatter and the mass trend of the richness--mass relation to other works. Subsequently, the impact of SZE contamination on our results is discussed. We then present other evidence for and against the contamination fraction and mean contaminant masses we have measured. Finally, alternative scenarios to the richness dependent contamination case are outlined, together with future prospects of discriminating among them.

\subsection{Comparison to literature}

Our comparison to the literature focuses on the two scaling relation parameters which are most closely linked to the richness dependent contamination:
the intrinsic scatter and the mass trend or power law slope. In the following, we consider also 
results based based on a redMaPPer selected sample obtained from Sloan Digital Sky Survey data \citep[SDSS, see ][for the discussion of the redMaPPer application]{rykoff14}, because those richnesses are consistent with the richnesses extracted from DES \citep[see][equations 66-67]{mcclintock19}. We will not include constraints based on the application of other optical cluster finders or richness measurement methods, as it is  unclear how they are impacted by projection effects compared to redMaPPer.

\subsubsection{Constraints on scatter}\label{sec:disc_scatter}
The intrinsic scatter in the observable-mass relation directly affects the mass incompleteness of cluster samples.  Measurements of the scatter are made even more important in many studies of 
optically-selected clusters, because weak lensing mass constraints are extracted from stacked observations of many clusters in bins of richness and redshift  \citep[e.g.][]{simet17, murata18, mcclintock19, murata19}. Due to the stacking, such studies lose 
leverage on the scatter, while Bayesian population modeling retains more of the constraining 
power in the data \citep{grandis19}. Noticeably, \citet{sereno20} recently suggested a Bayesian modelling approach to stacking that retains some constraining power on the intrinsic scatter. Cross calibration with ICM based mass proxies--- the technique applied in this work--- is a useful, and more widely used technique for constraining the scatter. 
\citet{rozo15} cross calibrated the SDSS-redMaPPer sample \citep{rykoff14} with the cluster catalog selected via SZE in the first all sky Planck survey \citep{planck13_clustercata}, investigating the scaling between the richness and the SZE-inferred masses of 191 clusters. The 
scatter around that relation was $\sigma_{\ln \lambda |M_\text{SZE}} = 0.266 \pm 0.017$. Investigating the relation between the 
DES-redMaPPer richness and the temperature in 58 archival \textit{Chandra} observations and 110 XMM observations, \citet{farahi19a} 
find $\sigma_{\ln\lambda |M} = 0.20^{+0.10}_{-0.08}$. Both of these measurements are in good agreement with our results, for 
instance $\sigma_\lambda = 0.22\pm0.06$.

\begin{figure}
	\includegraphics[width=\columnwidth]{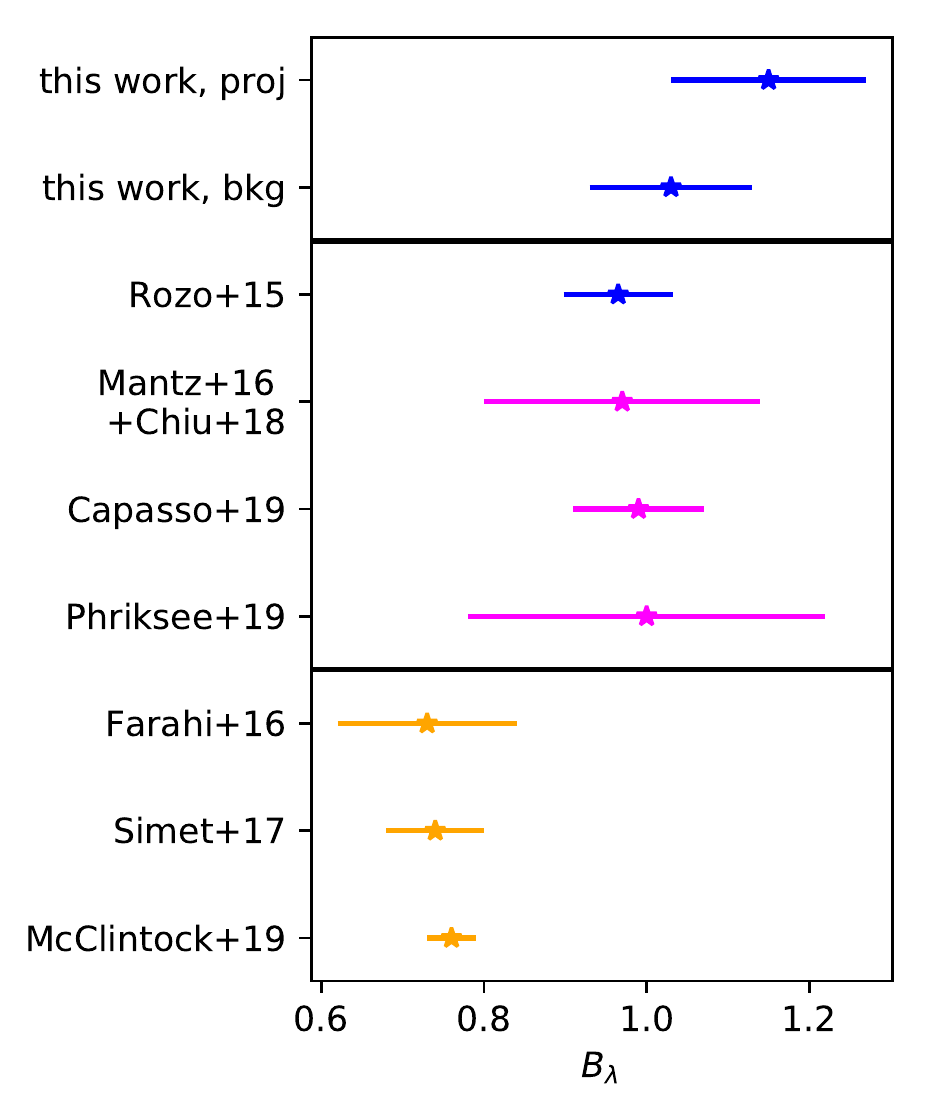}
	\vskip-0.10in
    \caption{Comparison of the constraints on the redMaPPer-richness--mass relation mass trend or power law
    slope from different works to our own results. The color 
    coding represents the selection method used: blue for SZE, magenta for X-ray and yellow for solely
    optical selection. The discrepancy between the results based on ICM selection (SZE and X-rays)
    and optical selection can clearly be seen. This difference may be due 
    to low richness sample contamination in analyses based on optically-selected samples.}
    \label{fig:discussion_blambda}
\end{figure}

\subsubsection{Mass trend of richness}

Our estimates of the mean contaminant mass is sourced mainly by the mis-match between our prediction for the mean mass in richness-redshift bins and its measurement through stacked weak 
lensing by \citet{mcclintock19}. Because the contamination fraction is larger at lower richnesses (see Fig.~\ref{fig:cont_of_lambda}), this contamination--- if ignored in a scaling relation analysis--- can manifest itself as an apparent difference in the mass trend parameter $B_\lambda$ (see also Fig.~\ref{fig:degens}).  Moreover, if different methods of cluster selection led to different levels of contamination, one might expect differences in the derived mass trend.

As an example, \citet{mcclintock19} finds a mass trend of $B_{\lambda, \text{McC19}} = 0.73\pm0.03$. The tension with our results is representative of the tension with various other results reported in recent years \citep[see][for summaries]{mcclintock19, capasso19b, bleem19}.  Examination of these results indicates a tendency for the mass trends in optically-selected samples to be shallower than those derived using samples that have ICM based selection.  


In the following and in Fig.~\ref{fig:discussion_blambda} we present a 
selection of those results. In the class of works based on ICM selection, we point the reader to
(1) \citet{rozo15} cross calibrating of SDSS-redMaPPer with clusters selected via SZE in the first all sky Planck survey that found $B_\lambda =0.965\pm0.067$;
(2) \citet{mantz16} determining $\lambda\sim M_\text{gas}^{0.75\pm0.12}$ on a sample ROSAT selected cluster, where combining this 
with the scaling $M_\text{gas}\sim M^{1.29\pm0.09}$ for the gas mass content of SPT selected clusters \citep{chiu18} results in 
$B_\lambda = 0.97\pm0.17$. Note here that this inference is based upon the relation between halo mass and gas mass. It is 
generally accepted from an empirical and theoretical perspective that the gas mass fraction of low mass systems is lower than that for high mass systems; see, e.g. \citet{mohr99} or the  
discussion in \citet[][section 5.2.2.]{bulbul19};
(3) a dynamical analysis of ROSAT selected, SDSS-redMaPPer confirmed cluster by \citet{capasso19b}, reports $B_\lambda = 0.99\pm0.08$;
and (4) a stacked weak lensing analysis of the same sample of ROSAT selected, SDSS-redMaPPer confirmed clusters by \citet{phriksee20}, reports $B_\lambda = 1.00\pm0.22$.

On the other hand, constraints obtained on purely optically-selected samples include:
(1) stacked spectroscopic analysis of SDSS-redMaPPer cluster by \citet{farahi16}, which finds $B_\lambda = 0.76\pm0.11$;
(2) stacked weak lensing of SDSS-redMaPPer clusters by \citet{simet17}, with $B_\lambda = 0.74\pm0.06$;
and (3) stacked weak lensing of DES-redMaPPer clusters by \citet{mcclintock19} with $B_\lambda = 0.73\pm0.03$.

Note that the stacked analysis of the CMB lensing signal around DES-redMaPPer clusters by \citet{baxter18} and \citet{raghunathan19a, raghunathan19b} currently does not constrain the 
mass trend of richness, and we do not include works that use number counts together with a fixed cosmology to constrain the mass slope \citep[e.g.,][]{murata18}, because the choice of cosmology can affect the mass slope. We also exclude from this discussion the study of the DES redMaPPer number counts and the large scale auto- and cross-correlations between galaxies, clusters and cosmic shear by \citet{to20}, as it is unclear if the mass information presented therein comes from the halo bias--mass relation, or from the combination of the number counts and the cosmological constraints from the auto- and cross-correlations.

Because the full cosmological analyses of the DES-redMaPPer sample \citep{des_y1_cluster} and the SDSS-redMaPPer sample \citep{costanzi19b} use the mass calibration derived by \citet{mcclintock19} and \citet{simet17}, respectively, we do not discuss them separately in this section.

Noticeably, constraints on ICM selected samples are in good agreement with our own results and generally indicate $B_\lambda\sim 1$ , whereas, results based in stacking methods of optically-selected samples suggest $B_\lambda\sim 0.75$, as can be clearly seen in Fig.~\ref{fig:discussion_blambda}. The redMaPPer sample contamination presented here could be the underlying driver of the apparent tension in these mass trend measurements. Specifically,  a contamination fraction which increases towards low richnesses would increase the $B_\lambda$ value inferred from the (stacked) WL data. The mass trend is of special importance to optical cluster cosmology; \citet{des_y1_cluster} identified this trend as the single most important source of the large tensions between redMaPPer based cluster cosmology constraints and those from other cosmological probes (including cosmological constraints from ICM selected samples). Specifically, a slope around 0.9 would make the number counts consistent with cosmological constraints from the shear and red galaxy auto- and cross-correlations \citep{desY1_3x2pt}.

\subsection{Impact of excess incompleteness in the SZE signal}

In this work, we implicitly did not explicitly include the impact of radio and dust emission on the SZE signal. We shall argue in the following why this assumption is justified and is not expected to change our inferred low mass contamination fraction. Two possible regimes of radio and dust emission need to be distinguished: 1) emission that is weak compared to the overall SZE signal of a halo, leading to an alteration of the SZE signal--mass relation, and 2) strong radio or dust emission, that wipes out the cluster SZE signature and leads to excess incompleteness in the SZE.

The first form of SZE contamination is implicitly included in this work. Our priors on the SZE scaling relation parameters have been derived empirically for the total SZE signal \citep{bocquet19}. Consequently, our priors include the impact of radio and dust emission, as long as these emissions do not suppress a majority of the halos SZE signal. The frequency of the latter has been studied by \citet{gupta17a} in the case of radio emission. In that work, the authors found that at the mass range we are interested in, the resulting SZE excess incompleteness is $<5\%$. That means that at most 5\% of the SPT-SZ cluster that should have been detected are lost due to radio emission. The implications on this work are that the predicted SPT confirmation fraction of redMaPPer objects $f(\text{SPT}|\hat\lambda, z) $ could at worst be 0.95 times the value that we computed assuming no excess SZE incompleteness from radio emission. Comparison to Fig.~\ref{fig:SPT_conf_fraction} shows that this is insufficient to describe the difference between the predicted and the measured confirmation fraction. Given that we interpret this difference as the presence of low mass contaminants, our estimation of the low mass contamination fraction is practically unaltered even by the worst case radio emission scenarios.

\subsection{Comments on the contamination}\label{sec:comment_on_cont}

Previously, we describe how our constraints on redMaPPer sample contamination provide a natural explanation for the tendency of analyses of optically-selected samples to lead to shallower mass trends.  Here we investigate the expected physical properties of the contaminants. Direct measurement of contamination in the richness range $20<\hat\lambda<40$ lies below the sensitivity of current wide area SZE based cluster surveys \citep[][]{bleem15}, although this is a regime opening up to improved RASS+DES samples such as MARD-Y3 \citep{klein19} and smaller but deeper SPT surveys \citep{huang19}.

However, the physical properties of the contaminants can be investigated using structure formation simulations. \citet{barnes17} studied the observable cluster properties in zoom-in hydrodynamical simulations of 30 galaxy cluster, using the \texttt{EAGLE} galaxy formation formalism. They found that galaxy groups of a mass $\sim3\times10^{13} M_\odot$ should have a soft X-ray ([0.5-2]keV band) luminosity of between $\sim2\times10^{42} \text{erg s}^{-1}$ and $\sim1\times10^{43} \text{erg s}^{-1}$ and ICM temperatures of $\sim1$ keV. Given sufficient X-ray sensitivity, we thus expect to see diffuse X-ray emission from the ICM of the contaminant systems. For instance, a follow up of SDSS-redMaPPer objects in the redshift range $0.08<z<0.12$ with $20<\hat\lambda<30$ by \citep{vonderlindenip} with the X-ray telescope \textit{Swift} will shed light on the luminosity distribution at low richnesses. Analysis of that data could potentially test our findings. Future prospects for detection with eROSITA are discussed below (cf. Section~\ref{sec:alternatives}).

According to the analysis of the number of satellite galaxies in halos performed by \citet{anbajagane20} on three suites of independent cosmological simulations of galaxy formation in a cosmological context (BAHAMAS + MACSIS, TNG300 of the IllustrisTNG suite, and Magneticum Pathfinder, all implementing different sub-grid physics prescriptions), our contaminants would have from 0 to 12 satellite galaxies above a stellar mass of $10^{10}$ M$_\odot$. It is unclear which fraction of those galaxies would appear as red sequence galaxy and thus be picked up by redMaPPer. Nevertheless, one would need to invoke projection effects to justify that a fraction of them might have richnesses of $>20$. This is however at odds with the projection effect model we employed in this work \citep{costanzi19} which has been calibrated for clusters with intrinsic richnesses as low as 5, and was set-up to account exactly for this effect. Given that we find that the projection effect model has very similar results compared to the background only model, this would imply that either the simulations or the projection model is inaccurate.
That is, either contaminants must have many more galaxies than expected for the mass we estimate or the projection effects must be much stronger.

 The projection model is derived assuming two main components: a redshift kernel that reflects the width of the red-sequence and a realisation of the distribution of red galaxy in an $N$-body simulation. The first component can be reliably derived from spectroscopic studies of cluster galaxies and the DES photometric properties. The latter was simulated by assigning richnesses to halos based on a richness--mass relation previously derived from stacked WL studies. A possible issue is that this relation might be biased by the presence of projection and selection effects. Another possible caveat is that the distribution of red galaxies as a function of local density in the simulation used to determine the projection model was not accurate outside of high mass halos. Such a mis-estimation of the density of red galaxies might bias the strength of projection effects. Empirical validation of the red galaxies distribution in dense environments predicted by the simulation, e.g. by comparing to measurements of the small scale bias of red galaxies, might improve our confidence in the accuracy of the projection model. Alternatively, direct empirical confirmation of the projection model is currently undertaken by direct spectroscopic studies of SDSS-redMaPPer objects in the redshift range $0.08<z<0.12$ \citep{mylesip}.

An effect that is degenerate with contamination as discussed in this work is the possibility that the weak lensing signal and the richness of clusters at the same mass are biased in the same (or opposite) direction by the same physical process \citep[see for instance][for simulation based work]{angulo12}. This would lead to correlated intrinsic scatter of the WL signal and of the richness (not unlike the correlated scatter among richness and SZE signal we discuss in Section~\ref{sec:correlated_scatter}). In the case of a correlation of the scatter between WL signal and richness physical, intuition suggests that mis-centering, halo triaxialty \citep{dietrich14} and project effects \citep{sunayama20} might lead to the correlation coefficient to be positive, as red galaxies trace dark matter tightly. This signal was already detected in simulations as `weak lensing selection bias' by \citet{des_y1_cluster}, but proved insufficient to account for the discrepancy in mass trends with respect to ICM based studies, as the correction did not show any strong richness trend. It is furthermore included in the mass estimates we used in this work, suggesting that it can not account for the signal that we interpret as contamination.

\subsection{Implications for redMaPPer cluster cosmology}

The implications of our result on the redMaPPer cluster cosmology which jointly fits for the redMaPPer number counts and stacked WL signals \citep[such as the analysis carried out by][]{des_y1_cluster}, deserve some further discussion. Firstly, our work implies that the richness--mass model used in that work, and derived by \citet{costanzi19}, is an incomplete description of the actual scatter between richness and mass. We investigate the possibility that the unresolved low-richness systematic advocated in \citet{des_y1_cluster} is related to such flawed modelling of the scaling relation. Specifically, we show that stronger projection effects are to be plausibly expected at low richness, as highlighted by our findings of a sizeable low mass population in the low richness range. We describe the possible reasons of this mis-calibration, and outline how the richness--mass modelling can be calibrated empirically (cf. Section~\ref{sec:comment_on_cont}).

Furthermore, we anticipate that using our constraints on the contamination fraction and mean contaminants mass as priors for a combined analysis of cluster abundance and stacked WL is likely to result in a strong deterioration of the cosmological constraints.
This is due to the fact that our inferred posteriors on the contamination fraction and mean contaminants masses are very broad. Especially allowing for the mean 
contaminants mass in each redshift-richness bin to vary within our posterior range is likely to strongly dilute the mass information gained from the stacked WL analysis, resulting in a weaker cosmological inference from the number counts. Nonetheless, we demonstrated in 
Section~\ref{sec:RM_numbercounts} that such weaker cosmological constraints will be consistent with the measurements for SPT-SZ number counts \citep{bocquet19} and the auto- and cross-correlations of galaxies and cosmic shear in DES-Y1 data \citep{desY1_3x2pt}.

\begin{figure}
	\includegraphics[width=\columnwidth]{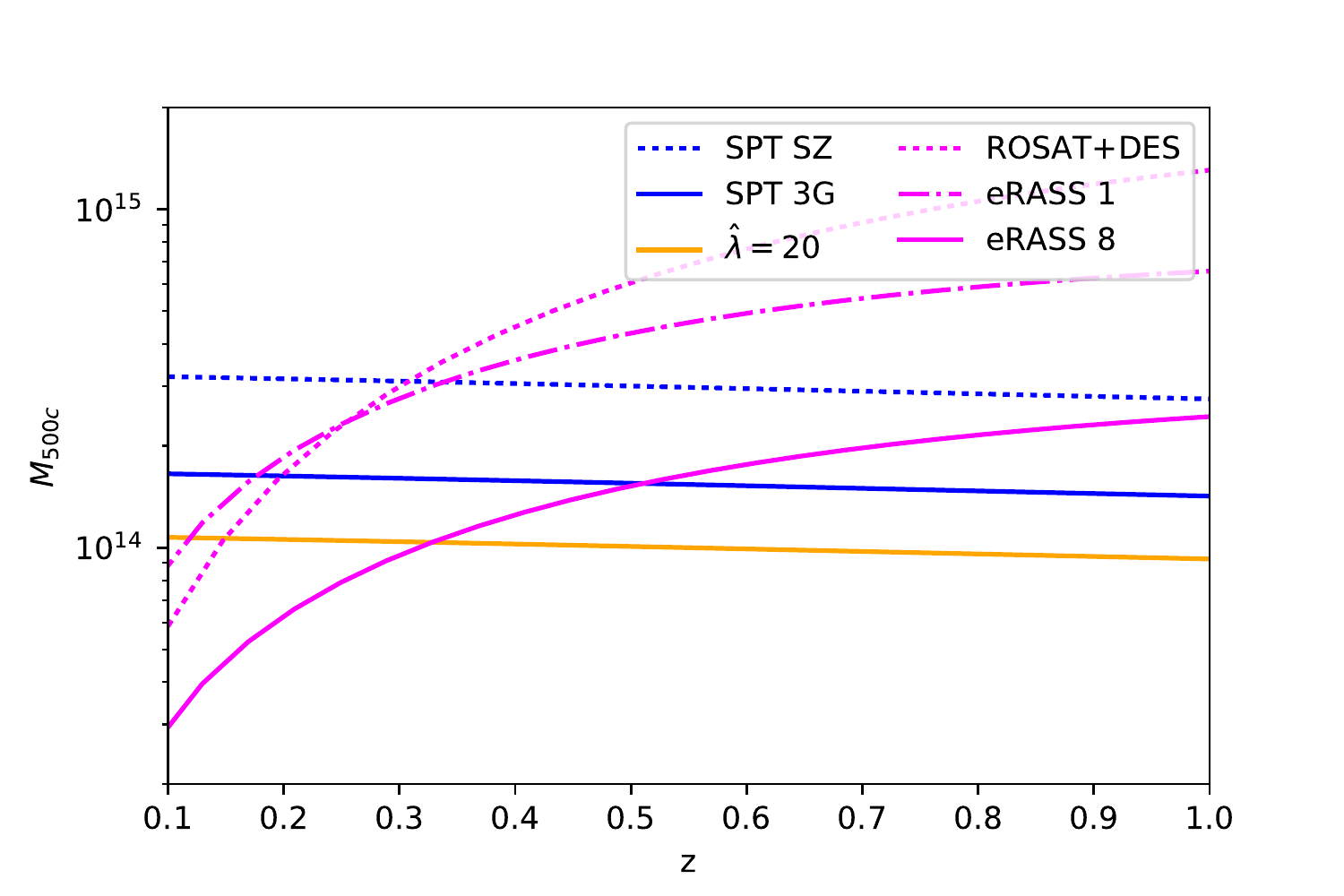}
	\vskip-0.10in
    \caption{Current (dotted lines) and future (solid and dashed-dotted) mass limits in 
    magenta for X-ray surveys and in blue 
    for SZE surveys. Yellow shows the mass corresponding to $\hat\lambda>20$. Deeper 
    photometric surveys do not lower the mass limit of 
    optical selection but enable detections to higher redshift. As can be seen in the 
    projected mass limits, eROSITA and 
    SPT-3G will extend the ICM leverage on low mass systems significantly.}
    \label{fig:prospects}
\end{figure}

\subsection{Alternative Explanations and Future Prospects}\label{sec:alternatives}

A possible question that we left unanswered is whether the low richness bins are actually populated by two populations with distinct mass distributions or rather by one single population with a significantly larger intrinsic scatter in mass. The latter option cannot be excluded from our analysis of the SPT-SZ confirmation fraction or the cross matched sample, because the SPT-SZ selection only selects the mid- and high-mass  portion of the mass distribution at any richness. This is precisely what the SPT-SZ confirmation fraction measures. Inspecting Fig.~\ref{fig:SPT_conf_fraction} one sees that less than a tenth of the objects with richness $\hat\lambda=40$ are detected. If the intrinsic scatter in richness were to increase strongly with decreasing richness or mass, this would not be visible in any of our observables, because we only observe the massive systems in the distribution of masses at richness $\hat\lambda=40$. The location of the mean of the distribution is thus degenerate with the width of the distribution (the manifestation in cluster studies of the Malmquist bias). As such, we cannot exclude that at low richnesses the population we labelled "clusters" and the population we labelled "contaminants" actually merge into a single population with large intrinsic scatter. Current constraints on the scatter (cf. Section~\ref{sec:disc_scatter}) are only applicable to the higher richness case.

This situation is however going to change in the next years as deeper SZE and X-ray surveys begin operations.  In the SZE regime, this can already be seen by the lower mass limit in the SPTpol survey \citep{huang19}. Further improvements are expected by the third generation SPT camera currently performing the SPT-3G survey over 1500 deg$^2$ \citep{benson14}. This will lower the limiting mass of SZE detection by a factor of 2 compared to SPT-SZ. In X-rays the recent "first light"\footnote{\url{http://www.mpe.mpg.de/7362095/news20191022}} of the 
eROSITA\footnote{\url{http://www.mpe.mpg.de/eROSITA}} X-ray telescope \citep{Predehl10,merloni12} on board the  Russian "Spectrum-Roentgen-Gamma" satellite will dramatically improve the sensitivity of X-ray cluster surveys. We present in Fig.~\ref{fig:prospects} the mass limits for 40 counts in the first eROSITA All-Sky-Survey (eRASS 1, 0.5 yr of observation) and eighth eROSITA All-Sky-Survey (eRASS 8, 4 yr of observation) following the prediction by \citet{grandis19}. This would allow us  to follow up 
optically-selected objects up to $z\sim0.35$ individually and to much higher redshifts in stacks of lower richness objects. The presence or absence of X-ray emission in eROSITA from optically 
selected objects will be a powerful tool to extend the kind of analysis presented in this work to lower richnesses. It will most 
likely help to discriminate the different scenarios we identified earlier.

\section{Conclusions} 

In this work we empirically validate the redMaPPer selected DES-Y1 survey by cross calibration with SPT-SZ selected clusters. We first limit ourselves to the high richness regime ($\hat\lambda>40$) to avoid optical incompleteness in the SPT confirmation. We produce a matched sample by positional matching between the redMaPPer-($\hat\lambda>40$) objects and the SPT-SZ selected clusters \citep{bleem15}. Of the 1005 redMaPPer selected cluster with measured richness $\hat\lambda>40$ in the joint footprint, 207 are confirmed by SPT-SZ. On the matched sample, we model the distribution in SZE-signal, richness and redshift with a 
Bayesian cluster population model. The free parameters of this model, the parameters controlling the scaling between richness and mass, and the scatter around this relation, are constrained from our analysis. We adopt  priors from previous SPT studies on the parameters on the SZE signal--mass relation, effectively transferring the vetted SPT mass calibration to the redMaPPer richness--mass relation. 

In an attempt to explore the impact of projection effects on the richness--mass relation, we employ two different error models: the first uses the error bars reported in the redMaPPer catalog and accounts only for the photometric uncertainty in the background subtraction, while the second includes projection effects calibrated on simulations \citep{costanzi19} . Our cross calibration of the richness--mass relation and the scatter around it is not significantly affected by the optical error model. Furthermore, our derived parameters are consistent with those reported by previous cross calibration studies \citep[e.g.][]{bleem19}.

We then turn to exploiting the information contained in the fact that some redMaPPer-($\hat\lambda>40$) in the joint SPT-DES Y1 footprint have been confirmed by SPT-SZ and some have not. Taking into account the relative SPT field depth at the redMaPPer positions, we employ the mass information contained in the richness and the SPT selection function to predict the detection probability by SPT for each redMaPPer cluster. We explore a model of the probability of SPT detection that accounts for possible redMaPPer contamination with respect to the scatter around the mean relation. Comparing these detection probabilities with the actual occurrence of matches constrains the contamination fraction.

Our investigation of the contamination fraction indicates that a model with a large contamination fraction of up to 50\% for richness 40 and a strong trend increasing to lower richness provides a better fit to the SPT confirmation fraction of redMaPPer clusters. Furthermore, the prediction of the redMaPPer number counts down to richnesses $\hat \lambda>20$ with a richness dependent contamination fraction are a better description of the measured number of clusters when compared to the case without contamination. This provides both internal and external evidence for considerable contamination, increasing to lower richnesses. Adopting our posterior on the contamination, our prediction for the clusters mean mass and the weak lensing constraints on mean mass in richness--redshift bins, we predict that the contaminants have mean masses of $\sim 3\times 10^{13} \text{M}_\odot$ ($\sim5\times10^{13}\text{M}_\odot$) for the range $20<\hat\lambda<30$ ($30<\hat\lambda<45$). The presence of group scale contaminants might be an explanation for the fact that in the cosmological study of redMaPPer objects \citep{des_y1_cluster} the WL mass measurements of richness-selected samples are biased low at low richness.

We discuss possible explanations for why analyses of the richnesses of ICM selected cluster samples tend to produce different (steeper) mass trends than analyses that rely on optically-selected cluster samples.  While this effect might indeed be due to the larger contamination of purely optically-selected samples, we also discuss that to date, the mass sensitivity of ICM selection does not extend to the mass range spanned by the richnesses  $20<\hat\lambda<40$.  As such, we can not exclude alternative explanations for the tension between our relation and the stacked weak lensing masses. We then highlight how upcoming X-ray surveys like eROSITA, and SZE surveys like SPT-3G will improve the ICM based selection sensitivities to probe the mass regime associated with these lower richness systems, enabling improved tests of the  results presented here.

\section*{Data Availability}

The data underlying this article is proprietary to the Dark Energy Survey Collaboration and the South Pole Telescope collaboration. Where applicable, please refer to the reference and links provided in this work to obtain the data.

\section*{Affiliations}

$^{1}$Faculty of Physics, Ludwig-Maximilians-Universit\"at, Scheinerstr. 1, 81679, Munich,  Germany\\
$^{2}$Excellence Cluster ORIGINS, Boltzmannstr. 2, 85748, Garching, Germany\\
$^{3}$Max Planck Institute for Extraterrestrial Physics, Giessenbachstr. 85748, Garching, Germany \\
$^{4}$INAF - Osservatorio Astronomico di Trieste, via G. B. Tiepolo 11, I-34143 Trieste, Italy\\
$^{5}$IFPU - Institute for Fundamental Physics of the Universe, Via Beirut 2, 34014 Trieste, Italy\\
$^{6}$Astronomy Unit, Department of Physics, University of Trieste, via Tiepolo 11, I-34131 Trieste, Italy\\
$^{7}$INFN - National Institute for Nuclear Physics, Via Valerio 2, I-34127 Trieste, Italy\\
$^{8}$Departamento de F\'isica Matem\'atica, Instituto de F\'isica, Universidade de S\~ao Paulo, CP 66318, S\~ao Paulo, SP, 05314-970, Brazil\\
$^{9}$Laborat\'orio Interinstitucional de e-Astronomia - LIneA, Rua Gal. Jos\'e Cristino 77, Rio de Janeiro, RJ - 20921-400, Brazil\\
$^{10}$Fermi National Accelerator Laboratory, P. O. Box 500, Batavia, IL 60510, USA\\
$^{11}$CEA, Physics Department, Durham University, South Road, Durham, DH1 3LE, UK\\
$^{12}$Institute of Cosmology and Gravitation, University of Portsmouth, Portsmouth, PO1 3FX, UK\\
$^{13}$CNRS, UMR 7095, Institut d'Astrophysique de Paris, F-75014, Paris, France\\
$^{14}$Sorbonne Universit\'es, UPMC Univ Paris 06, UMR 7095, Institut d'Astrophysique de Paris, F-75014, Paris, France\\
$^{15}$High Energy Physics Division, Argonne National Laboratory, 9700 South Cass Avenue, Lemont, IL 60439, USA\\
$^{16}$Kavli Institute for Cosmological Physics, University of Chicago, 5640 South Ellis Avenue, Chicago, IL 60637, USA\\
$^{17}$Department of Physics \& Astronomy, University College London, Gower Street, London, WC1E 6BT, UK\\
$^{18}$Kavli Institute for Particle Astrophysics \& Cosmology, P. O. Box 2450, Stanford University, Stanford, CA 94305, USA\\
$^{19}$SLAC National Accelerator Laboratory, Menlo Park, CA 94025, USA\\
$^{20}$Instituto de Astrofisica de Canarias, E-38205 La Laguna, Tenerife, Spain\\
$^{21}$Universidad de La Laguna, Dpto. Astrofísica, E-38206 La Laguna, Tenerife, Spain\\
$^{22}$Department of Astronomy, University of Illinois at Urbana-Champaign, 1002 W. Green Street, Urbana, IL 61801, USA\\
$^{23}$National Center for Supercomputing Applications, 1205 West Clark St., Urbana, IL 61801, USA\\
$^{24}$Institut de F\'{\i}sica d'Altes Energies (IFAE), The Barcelona Institute of Science and Technology, Campus UAB, 08193 Bellaterra (Barcelona) Spain\\
$^{25}$Institut d'Estudis Espacials de Catalunya (IEEC), 08034 Barcelona, Spain\\
$^{26}$Institute of Space Sciences (ICE, CSIC),  Campus UAB, Carrer de Can Magrans, s/n,  08193 Barcelona, Spain\\
$^{27}$Center for Cosmology and Astro-Particle Physics, The Ohio State University, Columbus, OH 43210, USA\\
$^{28}$Observat\'orio Nacional, Rua Gal. Jos\'e Cristino 77, Rio de Janeiro, RJ - 20921-400, Brazil\\
$^{29}$Centro de Investigaciones Energ\'eticas, Medioambientales y Tecnol\'ogicas (CIEMAT), Madrid, Spain\\
$^{30}$Department of Physics, IIT Hyderabad, Kandi, Telangana 502285, India\\
$^{31}$Department of Astronomy/Steward Observatory, University of Arizona, 933 North Cherry Avenue, Tucson, AZ 85721-0065, USA\\
$^{32}$Jet Propulsion Laboratory, California Institute of Technology, 4800 Oak Grove Dr., Pasadena, CA 91109, USA\\
$^{33}$Santa Cruz Institute for Particle Physics, Santa Cruz, CA 95064, USA\\
$^{34}$Institute of Theoretical Astrophysics, University of Oslo. P.O. Box 1029 Blindern, NO-0315 Oslo, Norway\\
$^{35}$Department of Physics and Astronomy, University of Missouri, 5110 Rockhill Road, Kansas City, MO 64110, USA\\
$^{36}$Instituto de Fisica Teorica UAM/CSIC, Universidad Autonoma de Madrid, 28049 Madrid, Spain\\
$^{37}$Department of Physics, Stanford University, 382 Via Pueblo Mall, Stanford, CA 94305, USA\\
$^{38}$School of Physics, University of Melbourne, Parkville, VIC 3010, Australia\\
$^{39}$School of Mathematics and Physics, University of Queensland,  Brisbane, QLD 4072, Australia\\
$^{40}$Department of Physics, The Ohio State University, Columbus, OH 43210, USA\\
$^{41}$Center for Astrophysics $\vert$ Harvard \& Smithsonian, 60 Garden Street, Cambridge, MA 02138, USA\\
$^{42}$Australian Astronomical Optics, Macquarie University, North Ryde, NSW 2113, Australia\\
$^{43}$Lowell Observatory, 1400 Mars Hill Rd, Flagstaff, AZ 86001, USA\\
$^{44}$Centre for Gravitational Astrophysics, College of Science, The Australian National University, ACT 2601, Australia\\
$^{45}$The Research School of Astronomy and Astrophysics, Australian National University, ACT 2601, Australia\\
$^{46}$Department of Physics and Astronomy, University of Pennsylvania, Philadelphia, PA 19104, USA\\
$^{47}$George P. and Cynthia Woods Mitchell Institute for Fundamental Physics and Astronomy, and Department of Physics and Astronomy, Texas A\&M University, College Station, TX 77843,  USA\\
$^{48}$Department of Astrophysical Sciences, Princeton University, Peyton Hall, Princeton, NJ 08544, USA\\
$^{49}$Instituci\'o Catalana de Recerca i Estudis Avan\c{c}ats, E-08010 Barcelona, Spain\\
$^{50}$Physics Department, 2320 Chamberlin Hall, University of Wisconsin-Madison, 1150 University Avenue Madison, WI  53706-1390\\
$^{51}$Institute of Astronomy, University of Cambridge, Madingley Road, Cambridge CB3 0HA, UK\\
$^{52}$Department of Physics and Astronomy, Pevensey Building, University of Sussex, Brighton, BN1 9QH, UK\\
$^{53}$School of Physics and Astronomy, University of Southampton,  Southampton, SO17 1BJ, UK\\
$^{54}$Computer Science and Mathematics Division, Oak Ridge National Laboratory, Oak Ridge, TN 37831\\
$^{55}$Department of Physics, University of Michigan, Ann Arbor, MI 48109, USA\\
$^{56}$Institute of Cosmology and Gravitation, University of Portsmouth, Portsmouth, PO1 3FX, UK\\
$^{57}$Department of Physics, The Ohio State University, Columbus, OH 43210, USA\\
$^{58}$Universit\"ats-Sternwarte, Fakult\"at f\"ur Physik, Ludwig- Maximilians Universit\"at M\"unchen, Scheinerstr. 1, 81679, M\"unchen, Germany

\section*{Acknowledgements}

We acknowledge financial support from the MPG Faculty Fellowship program, the DFG Cluster of Excellence 
``Origin and Structure of the Universe'', the new DFG cluster "Origins" and the Ludwig-Maximilians-Universit\"at Munich. AS is supported by the ERC-StG ‘ClustersXCosmo’ grant agreement 716762, by the FARE-MIUR grant 'ClustersXEuclid' R165SBKTMA, and by the INFN INDARK grant
Numerical computations in this work relied on the \texttt{python} packages \texttt{numpy} \citep{numpy} and 
\texttt{scipy} \citep{scipy}. The plots were produced using the package \texttt{matplotlib} \citep{matplotlib}. Posterior samples have been drawn from the likelihood functions and the priors using \texttt{emcee} \citep{emcee}. 

This work was performed in the context of the South-Pole Telescope scientific program.  SPT is supported by the  National  Science  Foundation  through  grant  PLR-1248097.  Partial support is also provided by the NSF Physics Frontier Center grant PHY-0114422 to the Kavli Institute  of  Cosmological  Physics  at  the  University  of Chicago,  the  Kavli  Foundation  and  the  Gordon  and Betty Moore Foundation grant GBMF 947 to the University of Chicago.  This work is also supported by the U.S.  Department  of  Energy.  AAS acknowledges support by US NSF AST-1814719.

This paper has gone through internal review by the DES collaboration. Funding for the DES Projects has been provided by the U.S. Department of Energy, the U.S. National Science Foundation, the Ministry of Science and Education of Spain, 
the Science and Technology Facilities Council of the United Kingdom, the Higher Education Funding Council for England, the National Center for Supercomputing 
Applications at the University of Illinois at Urbana-Champaign, the Kavli Institute of Cosmological Physics at the University of Chicago, 
the Center for Cosmology and Astro-Particle Physics at the Ohio State University,
the Mitchell Institute for Fundamental Physics and Astronomy at Texas A\&M University, Financiadora de Estudos e Projetos, 
Funda{\c c}{\~a}o Carlos Chagas Filho de Amparo {\`a} Pesquisa do Estado do Rio de Janeiro, Conselho Nacional de Desenvolvimento Cient{\'i}fico e Tecnol{\'o}gico and 
the Minist{\'e}rio da Ci{\^e}ncia, Tecnologia e Inova{\c c}{\~a}o, the Deutsche Forschungsgemeinschaft and the Collaborating Institutions in the Dark Energy Survey. 

The Collaborating Institutions are Argonne National Laboratory, the University of California at Santa Cruz, the University of Cambridge, Centro de Investigaciones Energ{\'e}ticas, 
Medioambientales y Tecnol{\'o}gicas-Madrid, the University of Chicago, University College London, the DES-Brazil Consortium, the University of Edinburgh, 
the Eidgen{\"o}ssische Technische Hochschule (ETH) Z{\"u}rich, 
Fermi National Accelerator Laboratory, the University of Illinois at Urbana-Champaign, the Institut de Ci{\`e}ncies de l'Espai (IEEC/CSIC), 
the Institut de F{\'i}sica d'Altes Energies, Lawrence Berkeley National Laboratory, the Ludwig-Maximilians Universit{\"a}t M{\"u}nchen and the associated Excellence Cluster Universe, 
the University of Michigan, the National Optical Astronomy Observatory, the University of Nottingham, The Ohio State University, the University of Pennsylvania, the University of Portsmouth, 
SLAC National Accelerator Laboratory, Stanford University, the University of Sussex, Texas A\&M University, and the OzDES Membership Consortium.

Based in part on observations at Cerro Tololo Inter-American Observatory, National Optical Astronomy Observatory, which is operated by the Association of 
Universities for Research in Astronomy (AURA) under a cooperative agreement with the National Science Foundation.

The DES data management system is supported by the National Science Foundation under Grant Numbers AST-1138766 and AST-1536171.
The DES participants from Spanish institutions are partially supported by MINECO under grants AYA2015-71825, ESP2015-66861, FPA2015-68048, SEV-2016-0588, SEV-2016-0597, and MDM-2015-0509, 
some of which include ERDF funds from the European Union. IFAE is partially funded by the CERCA program of the Generalitat de Catalunya.
Research leading to these results has received funding from the European Research
Council under the European Union's Seventh Framework Program (FP7/2007-2013) including ERC grant agreements 240672, 291329, and 306478.
We  acknowledge support from the Brazilian Instituto Nacional de Ci\^encia
e Tecnologia (INCT) e-Universe (CNPq grant 465376/2014-2).

This manuscript has been authored by Fermi Research Alliance, LLC under Contract No. DE-AC02-07CH11359 with the U.S. Department of Energy, Office of Science, Office of High Energy Physics.


\bibliographystyle{mnras}
\bibliography{manuscript} 


\appendix

\section{Optical completeness of the SPT sample}\label{sec:sptcompl}

\begin{figure}
	\includegraphics[width=\columnwidth]{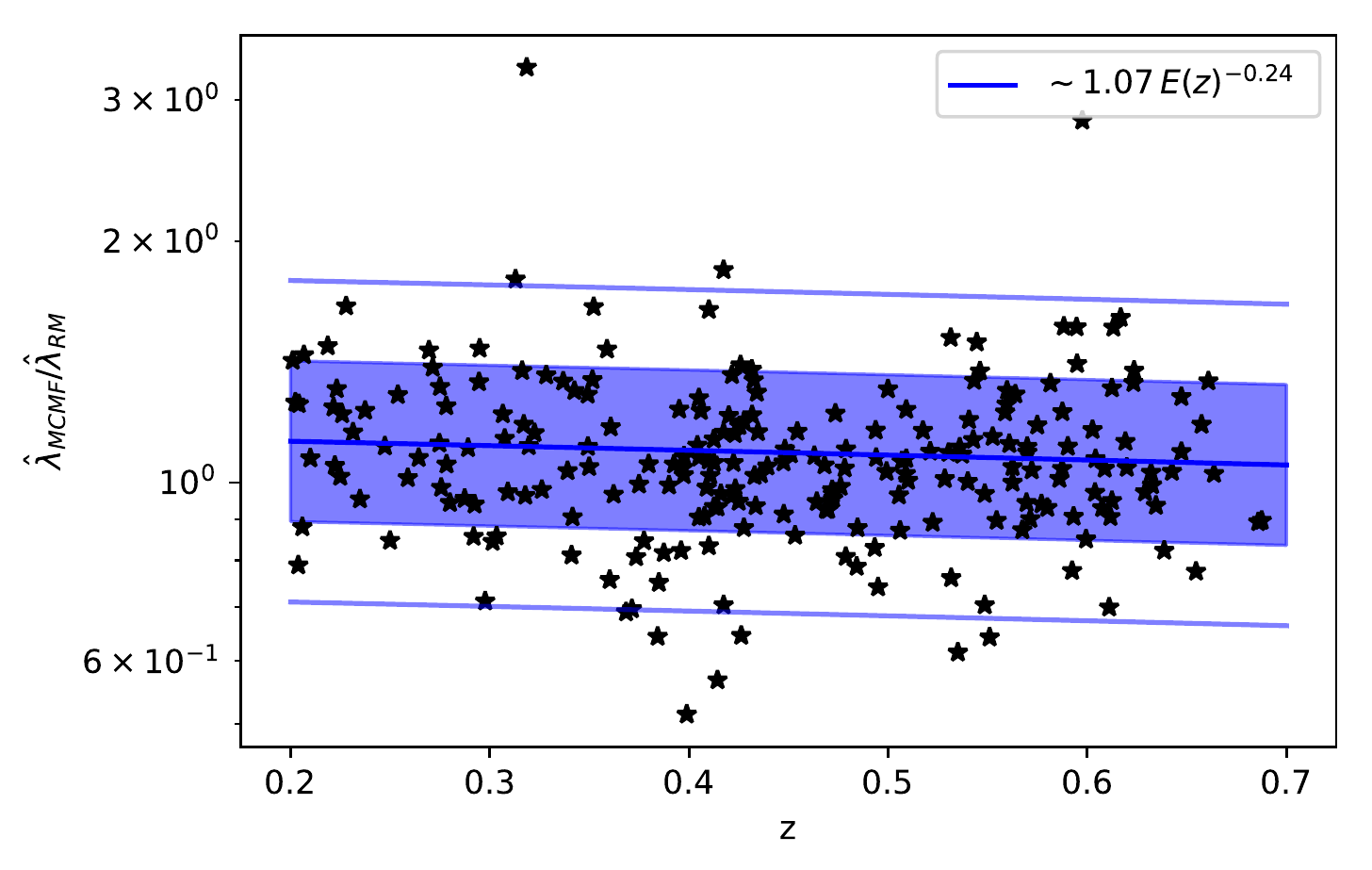}
	\vskip-0.10in
    \caption{Ratio between the MCMF richness $\hat\lambda_\text{MCMF}$ employed in the optical confirmation of SPT-SZ cluster to the redMaPPer richness $\hat\lambda_\text{RM}$ as a function redshift for the matched sample. In blue the mean relation between the two with the intrinsic scatter show by the shaded (1 sigma) and transparent lines (2 sigma). Different centering and apertures lead to intrinsic scatter, a deviation from unity and redshift trend.}
    \label{fig:lamMCMFoverlaRM_fit}
\end{figure}

As discussed above, we employ an SPT sample that has been confirmed with MCMF 
by imposing a cut $f_\text{cont}<0.1$, in the probability of random superpositions between the SZE candidate and physically unassociated optical structures.  As discussed previously 
\citep{klein19,kleinip, grandisip}, this cut is equivalent to a redshift dependent cut in the MCMF richness $\hat\lambda_\text{MCMF}>\lambda_\text{min}(z)$. 
Note that the richness $\hat\lambda_\text{MCMF}$ extracted by MCMF is not 
identical to the redMaPPer richness $\hat\lambda=\hat\lambda_\text{RM}$, 
because MCMF employs a prior from the SZE candidate on the position and the aperture. 
The ratio between the two richnesses on the matched sample is shown in 
Fig.~\ref{fig:lamMCMFoverlaRM_fit}. When fitting this relation we find
\begin{equation}
\frac{\hat\lambda_\text{MCMF}}{\hat\lambda_\text{RM}} = \big( 1.07 \pm 0.02 \big) \Big( \frac{E(z)}{E(0.6)} \Big)^{-0.24\pm0.02},
\end{equation}
with intrinsic log-normal scatter $0.23\pm0.01$.

Using this relation we estimate that objects with $\hat\lambda_\text{RM}>40$ are always more than 2$\sigma$ above the minimal 
MCMF richness for all redshifts we consider. This implies at least 97.5\% completeness. Every redMaPPer-($\hat\lambda>40$) object 
therefore (almost) certainly makes it past the SPT optical confirmation. In the case of the matched sample, where all objects 
have $\hat\lambda>40$, we therefore can safely ignore optical incompleteness in SPT. The same holds for the study of the SPT 
detection of redMaPPer-($\hat\lambda>40$), which we model as solely dependent on the SZE signal. We intentionally omit studying the 
redMaPPer detection probability of SPT objects. The probability of not finding an optically confirmed SPT cluster is given by the 
probability of $\hat\lambda_\text{MCMF}>\lambda_\text{min}(z)$ and $\hat\lambda<40$ at the clusters SZE signal and redshift, 
and thus depends on the optical incompleteness of the SPT sample.

\section{Cosmological dependence of the predicted redMaPPer number counts }\label{app:3x2pt}

\begin{figure}
	\includegraphics[width=\columnwidth]{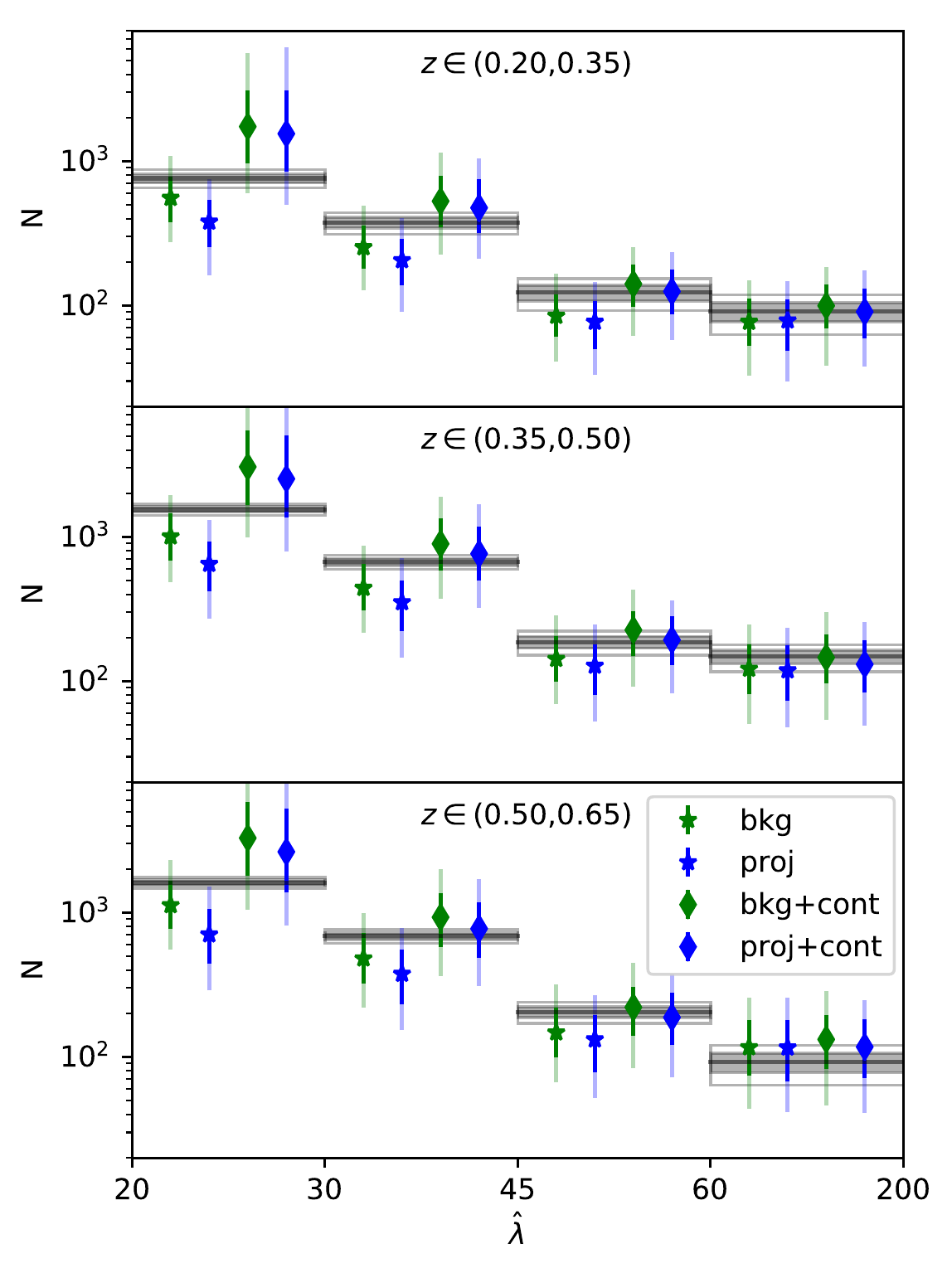}
	\vskip-0.10in
    \caption{The number of redMaPPer objects in bins of richness and redshift are shown as predictions (colored points) derived from our richness--mass relation 
    constraints and cosmology from the auto- and cross-correlations of cosmic shear and galaxies in the DES-Y1 data \citep{desY1_3x2pt} for the different optical error models (blue projection, green background). Stars denote predictions without contamination, while diamonds denote predictions with contamination. Over plotted in grey is the number of redMaPPer objects with the associated statistical uncertainty.  Also in this cosmology, the model with contamination fraction is not excluded by the number counts data.
    }
    \label{fig:3x2pt_comp_NC_marg}
\end{figure}

To test the cosmological sensitivity of agreement of our contamination fraction with the redMaPPer number counts, we predict the latter also assuming the cosmology derived from the auto- and cross-correlations of cosmic shear and galaxies in the DES-Y1 \citep{desY1_3x2pt}. Consequently, we used the priors $\Omega_\text{M} = 0.267 \pm 0.024$  and $S_8 = \sigma_8 (\Omega_\text{M}/0.3)^{0.2}=0.773\pm0.023$. The resulting predicted redMaPPer number counts are shown in Fig.~\ref{fig:3x2pt_comp_NC_marg}. Our qualitative assessment that the lower limit of the contamination fraction is consistent with the redMaPPer number counts holds also with these different cosmological priors.

\section{List of variables}

\begin{table*}
	\centering
	\caption{List of most relevant variable names used in this work with definition and with units where applicable.}
	\begin{tabular}{lll}\label{tab:vars}
		symbol & units & explanation \\
		\hline
		$M$ & $M_{\odot}/h$ &halo mass for spherical over-densities of 500 times the critical density of the universe \\
		$z$ &  & redshift (no distinction between true and measured redshift is made) \\
		$\lambda$ & & intrinsic richness \\
		$\hat \lambda$ & & measured richness \\
		$\zeta$ & & intrinsic SPT-SZ signal to noise \\
		$\xi$ & & measured SPT-SZ signal to noise \\
		$\gamma_f$ & & effective SPT-SZ field depth \\
		$\Omega_f$ & deg$^2$ & area of the SPT-SZ field $f$   \\
		$A_{\text{SZE},\lambda}$ & & amplitude of the SZE signal/richness -- mass relation \\ 
		$B_{\text{SZE},\lambda}$ & & power law index of the mass trend of the SZE signal/richness -- mass relation \\
		$C_{\text{SZE},\lambda}$ & & power law index of the redshift trend of the SZE signal/richness -- mass relation \\
		$\sigma_{\text{SZE},\lambda}$ & & intrinsic scatter around  the SZE signal/richness -- mass relation \\
		$\rho$ & & correlation coefficient between the intrinsic scatters around the SZE signal and richness -- mass relations \\
		$^{i}$ & & index for the $i$-th cluster (e.g. $\hat \lambda^{i}$ is the measured richness of the $i$-th cluster.) \\
		$N$ & & number of halos/clusters \\
		$\Omega_\text{M}$ & & present day fractional matter density of the Universe with respect to the critical density \\
		$\sigma_8$ & & root mean square of present day matter fluctuation amplitudes at a scale of 8 Mpc/h \\
		$\pi_\text{c}$ & & contamination fraction \\
		$^{j}$ & & index of the richness, redshift bin defined by the edges $(\hat\lambda_{-}^{j}, \hat\lambda_{+}^{j})$ and $(z_{-}^{j}, z_{+}^{j})$ \\
		$\bar M(j)$ & $M_{\odot}/h$ & mean mass of the bin $j$ \\
		$\hat M_\text{WL}^{j}$ & $M_{\odot}/h$ & measured mean WL mass in the bin $j$ \\
		$\delta \hat M_\text{WL}^{j}$ & $M_{\odot}/h$ & measurement uncertainty on mean WL mass in the bin $j$ \\
		$\bar M_\text{c}^{j}$ & $M_{\odot}/h$ & mean contaminants mass of the bin $j$ \\
	\end{tabular}
\end{table*}

In table~\ref{tab:vars} we summarize the variables used in this work with units and explanation.


\bsp	
\label{lastpage}
\end{document}